\documentclass[epj,nopacs,final]{svjour}
\usepackage[ansinew]{inputenc}
\usepackage{lmodern}					% Ersatz fuer Computer Modern-Schriften
\usepackage[T1]{fontenc}	

\usepackage{geometry}
\usepackage{placeins}
\usepackage{subcaption}
\usepackage{amsmath}
\usepackage{amsfonts}
\usepackage{xcolor} 
\usepackage{graphicx}
\usepackage{float}
\usepackage{afterpage}
\usepackage{wrapfig}
\usepackage{bm} 							% bold math
\usepackage{bbold} 							% for mathbb
\usepackage[
pdftitle={Tri- and multiple bistable activity in random binary networks},
pdfauthor={Christ, Simon; Sonnenschein, Bernard; Schimansky-Geier, Lutz},
pdfkeywords={ two-state, heterogeneous mean-field approximation, complex networks },
pdfpagemode=UseOutlines,
colorlinks=true,  
linkcolor=black,   
filecolor=black,    
urlcolor=black,      
citecolor=black,      
pdftex=true,           
plainpages=false,       
hypertexnames=false,     
pdfpagelabels=true,       
hyperindex=true
]{hyperref}

\newcommand{\Pvec}{\mathcal{P}_1}
\newcommand{\PbetI}{P^*(1|k_1)}
\newcommand{\PbetII}{P^*(1|k_2)}
\newcommand{\gI}{\gamma_{k_1}}
\newcommand{\gII}{\gamma_{k_2}}

\newcommand{\adj}{\text{adj}}

\begin{document}
\title{Tristable and multiple bistable activity in complex random
  binary networks of two-state units}
%\subtitle{Do you have a subtitle?\\ If so, write it here}
%
\titlerunning{Tri- and multiple bistable activity in random binary
  networks} 
\authorrunning{Simon Christ et al.}  
\author{Simon Christ \thanks{Present address: Theory \& Bio-Systems
    Department, Max-Planck-Institute of Colloids and Interfaces,
    D-14424 Potsdam-Golm, Germany} \and Bernard Sonnenschein \and Lutz
  Schimansky-Geier}
%
%\offprints{}          % Insert a name or remove this line
%
\institute{ Department of Physics, Humboldt-University at Berlin,
  Newtonstr. 15, D-12489 Berlin, Germany;\linebreak
  \email{Simon.Christ@mpikg.mpg.de, sonne@physik.hu-berlin.de,
    alsg@physik.hu-berlin.de} }
\date{Received: date / Revised version: date}
% The correct dates will be entered by Springer
%
\abstract{ We study complex networks of stochastic two-state units.
  Our aim is to model discrete stochastic excitable dynamics with a
  rest and an excited state.  Both states are assumed to possess
  different waiting time distributions.  The rest state is treated as
  an activation process with an exponentially distributed life time,
  whereas the latter in the excited state shall have a constant mean
  which may originate from any distribution.  The activation rate of
  any single unit is determined by its neighbors according to a random
  complex network structure.  In order to treat this problem in an
  analytical way, we use a heterogeneous mean-field approximation
  yielding a set of equations general valid for uncorrelated random
  networks.  Based on this derivation we focus on random binary
  networks where the network is solely comprised of nodes with either
  of two degrees. The ratio between the two degrees is shown to be a
  crucial parameter.  Dependent on the composition of the network the
  steady states show the usual transition from disorder to homogeneous
  ordered bistability as well as new scenarios that include
  inhomogeneous ordered and disordered bistability as well as
  tristability. The various steady states differ in their
    spiking activity expressed by a state dependent spiking rate.
  Numerical simulations agree with analytic results of the
  heterogeneous mean-field approximation.
}%end of abstract

\maketitle
%

%\onecolumn
\section{Introduction}
\label{intro}
Discrete-state stochastic models can be used to describe discrete
processes such as the orientation of a spin or the blinking of quantum
dots \cite{Frantsuzov09}. Additionally, systems with continuously
changing dynamics can be mapped to discrete-state descriptions via
coarse-graining \cite{Lindner2004,HuberTsim2005,PikovskyTsim2001}.
Despite the simplicity of the single units, the collective effects of
ensembles of coupled units can be highly non-trivial.

In earlier works discrete stochastic three-state models have been used
to investigate fluctuation driven spin nucleation on complex networks
\cite{Chen15}. Two- and three-state models have been applied to
neuronal systems \cite{Dumont2014,PrFaSchiZa07} and recently to
language dynamics \cite{Colaiori15}. Synchronization behavior, phase
transitions and reaction to time delayed feedback
\cite{WoodLind06,WoodLind07,WoodLind06PRE,Prager2003,EscaffLind12,Kouvaris2010,HuberTsim2005,PikovskyTsim2001}
as well as excitability \cite{Prager2007,Leonhardt2008} are general
aspects that apply to a wide range of natural phenomena.

Most of the referenced works consider Markovian -- thereby memoryless
-- discrete-state models
\cite{WoodLind06,WoodLind07,WoodLind06PRE,Pinto2014}. A disordered
environment \cite{Trimper04} or the reduction of models with a high
number of discrete states to a model with fewer states generically
demands a non-Markovian description. Therefore, in continuation of
previous work \cite{Prager2003,PrFaSchiZa07,Prager2007} in this paper
a semi-Markovian model \cite{cox1962renewal} of stochastic two-state
units is considered.  As a new point of interest these units are
embedded in an uncorrelated random network \linebreak whose nodes
possess different but independent degrees, in particular this excludes
networks with high asortativity or dissortativity.  The structure of
the network is given by the node distribution $p(k)$. A big number of
nodes is assumed, it is known that finite size effects
\cite{Pinto2014} have a strong effect on the dynamic of the stochastic
process as well as on the network influence.  Complex networks are of
top interest in statistical physics because it allows deviation from
global coupling without specification of a spatial structure as well
as providing a framework to map complex spatial structures to an
abstract space.  Furthermore, network structures are present in many
situations of everyday life, for example transport
\cite{Barrat2004,Jiang2007,Mukherjee2003} and supply networks
\cite{Runions2005} to mention just a few.

The paper is structured as follows.  In Section \ref{sec:MF} the
master equations for sto\-chastic two-state units are derived. In
Section \ref{sec:HMF} the heterogeneous mean-field approximation is
used to reduce the set of equations. Section \ref{sec:RBN} applies the
formalism to random binary networks and consists of three subsections.
In the first subsection the implicit equations for the steady states
are derived and analyzed for saddle-node bifurcations.  From this, the
scaling of the critical coupling strength due to the network embedding
is revealed. The second subsection addresses the limit of vanishing
noise.  Finally, the last subsection shows the results of the
numerical solutions of the mean-field equations in presence of finite
noise.  Apart from the expected homogeneous ordered bistability that
is known from globally coupled units \cite{Prager2003,PrFaSchiZa07},
inhomogeneous ordered and disordered states as well as tristable
states are uncovered.  These findings are confirmed via microscopic
simulations using the original network structure. In these
  simulations the network exhibits a variable firing activity by
  approaching different steady states in dependence on the initial
  conditions. 
  But the firing of the populations is not synchronized and thus the mean-field stays constant. 
  The firing differs in the spike generating
  rates for the two populations and whether these are in the rest or
  excited states. The spike time statistics of
  individual units are discussed separately. 
  Given that the excited state possess an
  exponentially distributed waiting time, the spike trains are always
  nearly Poissonian. 
  For a sharp-peaked waiting time density with no variance,
  the spike trains become highly coherent.

\section{Two-state processes on complex networks}
\label{sec:MF}
\subsection{Master equation of coupled two-state units}
\label{sec:ME2}
\begin{figure}[!htb]
  \centering
  \includegraphics[width=.49\textwidth]{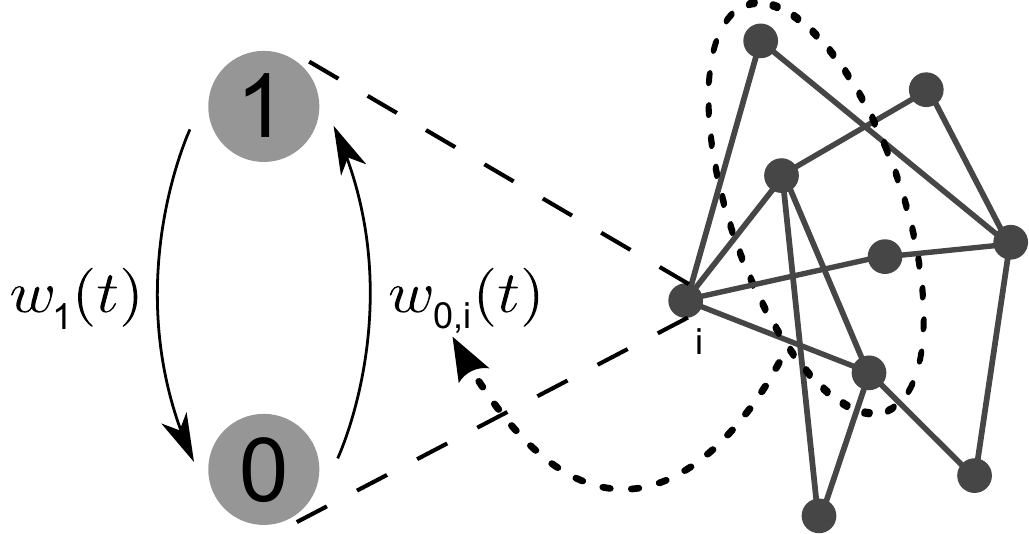}
  \caption{Sketch of the considered units and their topology. The
    two-state units with the waiting-time distributions $w_{0,i}(t)$
    and $w_1(t)$ are embedded in a binary random network, as indicated
    by the dashed lines.  In this example are five nodes with degree
    two and four nodes with degree four.  The highlighted unit $i$ is
    one of the latter, hence its activation waiting-time distribution
    $w_{0,i}(t)$ is affected by the activity of four neighbors, as
    indicated by the dotted ellipse and the dotted arrow.  }
\label{fig:TSO}
\end{figure}

The stochastic two-state units considered in this work can switch
between the states $0$ and $1$. They do so in a stochastic fashion,
governed by the waiting time distributions $w_{0,i}(t)$ and $w_1(t)$
in the corresponding states, as explained Fig.~\ref{fig:TSO}. Focusing
on excitable dynamics, state $0$ will be called the resting state and
state $1$ the excited state.
In this way, the transition from rest to excited, i.e. from state $0$
to state $1$, will be assumed as an activation process.  The life time
in state $0$ is exponentially distributed and the state will be left
with the transition rate $\gamma$.

The backward transition (relaxation from excited to rest) is governed
by the waiting time density $w_1(t)$ to remain in state $1$.  So far
this density is arbitrary except that its mean value shall be $\tau$
and does not depend on any other parameters such as the noise
intensity, the size and structure or possible dynamical states of the
network.  In the simulations two specific choi\-ces are made.  First,
the relaxation is treated as a Markovian rate process with a
exponential waiting time distribution
\begin{equation}	
w_1(t)=\frac{1}{\tau}\exp\left(-\frac{t}{\tau}\right).
\label{eq:w1}
\end{equation}
Oppositely, when modeling excitable dynamics, a sharply peaked
distribution is more suitable, e.g.
\begin{align}
	w_1(t)=\delta(t-\tau)
	\label{eq:w1peak}
\end{align}
would yield a constant waiting time without any variance in state $1$
modeling a fixed delay \cite{Prager2003,Prager2007,Kouvaris2010}.  As
shown in the appendix~\ref{sec:app}, the bifurcations of the steady
states are not affected by the specific choice of $w_1$. It depends on
the mean time $\tau$ spent in the excited state. Only, setups with
possible transition to a non-stationary behavior reflect on the choice
of $w_1$.

The two-state units are located on the $N$ nodes of a complex
network. Each stochastic element receives the output from the units
to which it is linked in the network by edges.  This coupling
is mathematically realized by introduction of the adjacency matrix
$\tens{A}$. The activation rate $\gamma_i$ from state $0$ to state $1$
of the $i$th node $i\in\{1,\dots,N\}$ is assumed to depend on a signal
function $f_i(t)$
\begin{align}
	\gamma_i &= \gamma[f_i(t)],
\end{align}
which contains the adjacency matrix
\begin{align}
	f_i(t) &= \frac{1}{N}\,\sum_{j=1}^N A_{ij}\, s_j(t).
	\label{eq:output1}
\end{align}
Therein, $s_j(t)$ is the output signal of node $j$ depending on its
state.  In this work, undirected and unweighted networks are
considered, hence the adjacency matrix is symmetric with elements
$A_{ij}=1$, if the units $i$ and $j$ are connected, otherwise
$A_{ij}=0$.

Typically, excitable systems stay in the resting state where no output
is produced. Upon sufficient excitation they drastically change their
intrinsic dynamics which is emitted as a signal.  Here such
``spiking'' is modeled by a two-valued output function $s_j(t)$, which
can take the values $1$ in the excited state and $0$ otherwise.  This
setting is motivated by neuronal activity or excitable lasers. A
symmetric choice would be more appropriate in order to model magnetic
spins.

Eventually, in accordance with previous assumptions, the normalized
waiting time distribution density of the activation with the time
dependent rate from \eqref{eq:output1} is given by the exponential
function
\begin{equation}
\label{eq:w0}
  w_{0,i}(t) = \gamma[f_i(t)] \,\exp\left(-\int_0^t\,\gamma[f_i(t')] \,
  \mathrm{d} t' \right).
\end{equation}

Let $P_{i}(0,t)$ denote the occupation probability of unit $i$ to be in state $s_i(t)=0$ at time $t$. 
Analogously $P_{i}(1, t)$ for state $s_i(t)=1$. 
Then the balance of probability flows yields the generalized master equations
\begin{equation}
	\begin{aligned}
		\dot{P}_i(0,t)&=-J_{0\rightarrow
                  1,i}(t)+J_{1\rightarrow
                  0,i}(t),\\ \dot{P}_i(1,t)&=-J_{1\rightarrow
                  0,i}(t)+J_{0\rightarrow 1,i}(t),
	\end{aligned}
	\label{eq:flows}
\end{equation}
for all $i\in\{ 1, \dots, N \}$, where $J_{0\rightarrow 1,i}(t)$ gives
the probability flow from state $0$ to $1$ of unit $i$ at time
$t$. Since the transition $0\rightarrow 1$ is a rate process, its
probability flow is simply given by
\begin{equation}
	J_{0\rightarrow1,i}(t) = \gamma[f_i(t)]P_i(0,t).
\end{equation}
The second probability flow is given by all the probability that has
flown into state $1$ up to time $t$ and stayed there for a time, which
is given by the waiting time distribution $w_1(t)$.  Thus it is the
convolution of $J_{0\rightarrow1,i}(t)$ and $w_1(t')$,
\begin{equation}
	J_{1\rightarrow0,i}(t) = \int_0^{\infty}\gamma[f_i(t-t')]P_i(0,t-t')w_1(t')\mathrm{d} t'.
\end{equation}
Using the normalization condition at a given node
\begin{equation}
	P_i(0,t)=1-P_i(1,t),
	\label{eq:norm_i}
\end{equation}
the temporal evolution of the occupation probabilities $P_i(1,t)$ is
then given by \linebreak (cf.~\eqref{eq:flows})
\begin{align}
	\nonumber &\dot{P}_i(1,t) = \gamma[f_i(t)](1-P_i(1,t))- \\
	\nonumber &-\int_0^{\infty}\gamma[f_i(t-t')](1-P_i(1,t-t'))w_1(t') \mathrm{d} t', \\
	& \hspace{3.7cm} i \in \{ 1, \dots, N \}.
	\label{eq:dtP2}
\end{align}

This is a set of $N$ coupled linear integro-differential
  equations for the $P_i(1,t)$. The complexity is given by the
  chosen adjacency matrix $A_{i,j}$ which links the almost $N$
  equations by the the signal function, see Eq. \eqref{eq:output1}.
%Note the importance of $w_0(t)$ to be
%exponential.  This justifies to account for one cycle only, otherwise
%it would in principle be necessary to trace back all cycles from point
%$t$ to the beginning.  

By applying a heterogeneous mean-field approximation as described in Section \ref{sec:HMF}, the structure of this set of equations is
much simpler and the number of differential equations in the set
reduces significantly. As consequence of this approximation, the set
will become analytically treatable for special cases in the stationary
limit.

But before describing this approximation, an
appropriate notation for the values describing the dynamical
behavior on a complex network will be introduced.

\subsection{Master equation on complex networks}
\label{sec:ME}
To include the network topology into the description, it is assumed
that the complex network structure can be described as a random
network.  Central value in this description are the degrees $k_i$ of
the $N$ nodes.  For a given adjacency matrix $A_{i,j}$ with values
$0,1$ the degree $k_i$ of the $i$th node is the number of existing
links to other nodes of the network. It becomes
\begin{align}
    \label{eq:degree}
            k_i = \sum_{j=1}^{N} A_{i,j}.
\end{align}

Let $k_{\mathrm{min}}$ and $k_{\mathrm{max}}$ be the minimum and
maximum degree occurring in the network, respectively, while $N_k$ is
the number of units with degree
$k\in[k_{\mathrm{min}},k_{\mathrm{max}}]$.  In a random network the
degrees can be treated as random numbers.  Their occupation
probability $p(k)$ is defined by $N_k$.  Then the occupation
probability reads
\begin{align}
    \label{eq:degree_occ}
            p(k) =  \lim_{N\to \infty }\frac{N_k}{N}.
\end{align}
Mathematically, the limit make sense if $N_k~\propto~N$.
  Afterwards, sorting the nodes with coinciding degrees $k$ in
  the network gives the joint probability that any node in the
  network has this degree and is, respectively, in the dynamical state
  $s=(0,1)$ at time $t$
\begin{align}
    \label{eq:degree_prob}
          P(s,k,t)  = \sum_{i=1}^{N}  \,\delta_{s,s_i}\, \delta_{k,k_i}\,P_i(s_i,k_i,t).
\end{align}
Using Bayes' theorem we split off the occupation probability $p(k)$ as
\begin{align}
    \label{eq:Bayes}
          P(s,k,t)  =  P(s,t|k)p(k).
\end{align}
Normalization reads
\begin{align}
    \label{eq:norm}
          P(0,k,t) \,+\,P(1,k,t)  =  p(k).
\end{align}
Hence, for the conditioned probabilities the degree is fixed and 
  it becomes
\begin{align}
    \label{eq:norm_c}
         P(0,t|k)\,+\, P(1,t|k) = 1.
\end{align}

The chosen conditional probabilities $P(1,t|k)$ neglect that various
nodes with the same degree $k$ might be linked to $k$ nodes with
different degrees. Therefore, without a further approximation, they
are so far not suitable for the description of our situation.

To proceed, the signal function \eqref{eq:output1} which is
coupling the $i$th node to the other nodes will be considered now.
Replacing therein the number of the linked node $j$ by the specific
degree $k_j$ of this node, it becomes
\begin{align}
    \label{eq:signal_degree}
         f_i(t) =  \frac{1}{N}\, \sum_{j=1}^{N}\,A_{i,j}s_{k_j}(t).
\end{align}
Sorting different degrees gives 
\begin{align}
    \label{eq:signal_degree1}
         f_i(t) =  \sum_{k=k_{min}}^{k_{max}}\, \frac{N_k}{N} \, \frac{1}{N_k}\,\sum_{j=1}^{N}\,\delta_{k,k_j}\, A_{i,j}\,s_{k_j}(t).
\end{align}
Every node receives its input depending on the specific linked
environment. Due to the disorder contained in the adjacency matrix $A_
{i,j}$, the summation is still in general different for distinct
nodes. The reduced description for given degrees requires an
additional approximation.

% Assuming that
%   $N_k,\propto N$ and that there is no preference between the degree
%   of node $i$, one might formulate in the limit of a large number $N$
% }
% \begin{align}
%     \label{eq:signal_degree2}
%          \lim_{N \to \infty }f_i(t) =  \sum_{k=k_{min}}^{k_{max}}\, p(k) \,
%          \langle A_{i,j} \rangle_{s_{k_{j}}=1,k_j=k}.
% \end{align}
% Consequently, the r.h.s. does not contain correlations between
%   the degrees of the nodes. }

\subsection{Heterogeneous mean-field approximation}
\label{sec:HMF}
\begin{figure}[H]
	\centering
        \includegraphics[width=\linewidth]{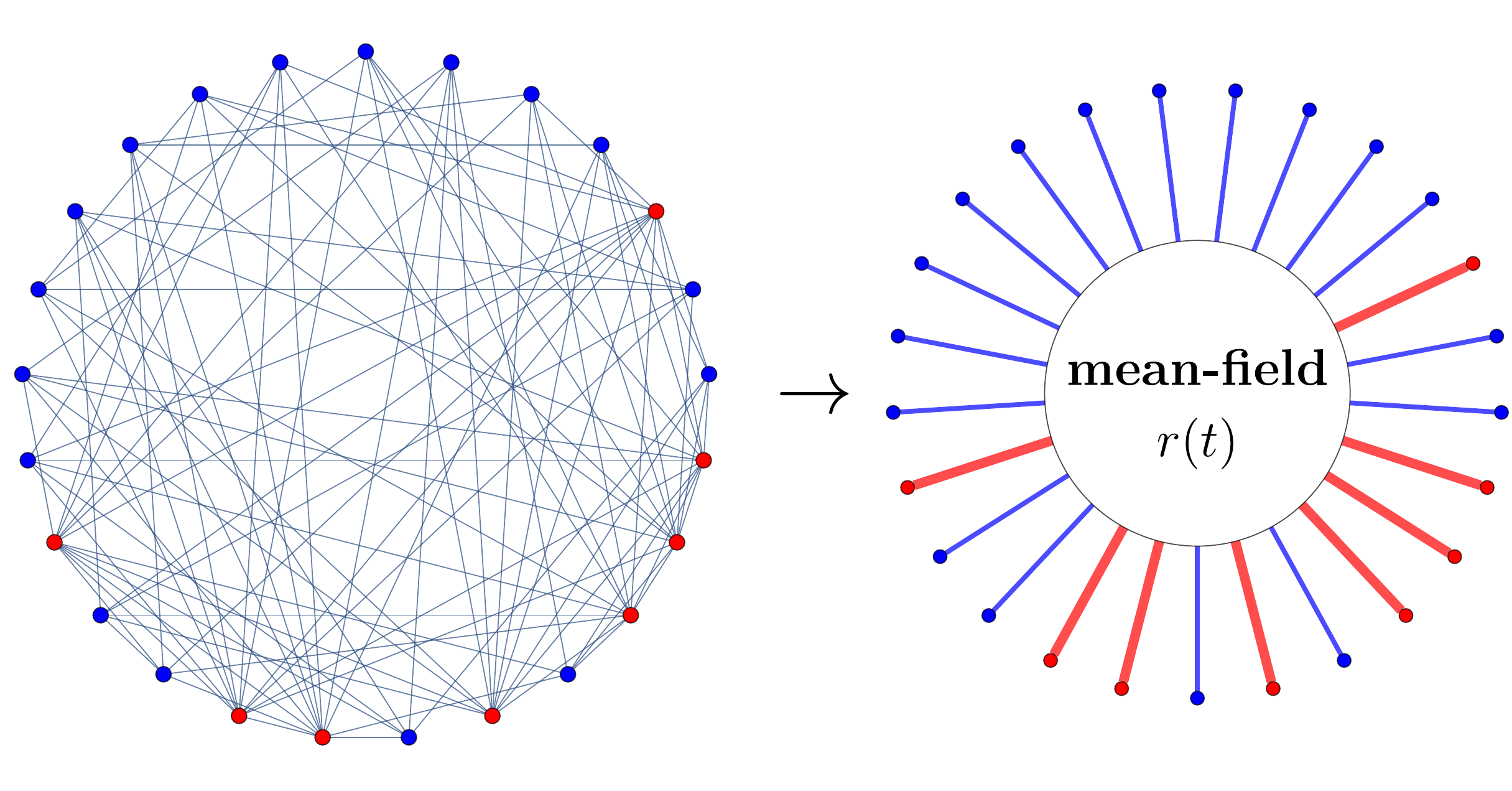}
	\caption{Representation of a random binary network with
          $k_1=12$, $k_2=6$ and $N=25$ before and after applying the
          heterogeneous mean-field approximation.}
	\label{fig:HMF}
\end{figure}

As proposed in \cite{Ichinomiya2004,Sonnenschein2012,Carro2016}, the
complex network with the adjacency matrix $A_{i,j}$ and with the given
degree distribution $p(k)$ will be replaced by a fully connected
network with weighted edges. The latter with adjacency matrix
$\tilde{A}_{i,j}$ shall possess the same distribution of the degrees
at the nodes. In detail, it is required that the degree-values of all
nodes in the new network shall coincide with the corresponding ones in
the original network.  Accordingly, the following shall hold
\begin{align}
    \label{eq:degree1}
    \sum_{j=1}^{N} A_{i,j}\, = \,k_i\,= \sum_{j=1}^{N} \tilde{A}_{i,j}.
    \quad i=1,\ldots,N.
\end{align}
Whereas the sum at the l.h.s. is running over values $1$ and $0$ of
$A_{i,j}$, at the r.h.s. the sum goes over rational numbers.
The assumption is made that the probability to have an edge between
the $i$th and $j$th node is proportional to the product of the degrees
of these nodes, i.e., to $k_i\,k_j$. This is strictly valid only for
uncorrelated networks.  Further on, taking into account the
conservation of degree as required in \eqref{eq:degree1} the following
replacing for a fully connected network is defined
\begin{equation}
  \tilde{A}_{ij} =  \frac{ k_i\, k_j}{\sum_{l=1}^{N} k_{l}}.
  \label{eq:newA}
\end{equation}
Figure~\ref{fig:HMF} illustrates the replacement. As a result the
single nodes with given degree $k$ couple uniquely to a mean filed.
Hence, in contrast to the original network, the fully connected
network will allow a mean-field representation with respect to all
edges with coinciding degrees $k$. In \cite{Sonnenschein2012} the
validity of this replacement procedure was discussed and it was
successfully applied to complex networks with continuous phase
oscillators at the nodes. The approach
\cite{Ichinomiya2004,Sonnenschein2012,Carro2016} is here generalized
to discrete stochastic two state units.

%%%%%%%%%%%%%%%%%%%%XS

As a result of the replacement, the signal function in
\eqref{eq:signal_degree1} is approximated as
\begin{equation}
  \tilde{f}_i(t)\approx\frac{1}{N}\sum_{j=1}^N \tilde{A}_{ij}s_j(t)=\frac{k_i}{N\sum_{l=1}^{N}k_{l}}\sum_{j=1}^{N}k_j s_j(t).
  \label{eq:newoutput}
\end{equation}
It follows immediately, that the signal at the node with degree $k_i$
coincides for all nodes with the same degree.  The node value itself
enters only multiplicatively into this expression.  Hence, the
denominator of the sum is equal for all nodes.

Crossing from the summation over all nodes to sums with the same
  degrees similar transformations as above are made. The different
  degrees $k_i$ are again divided into classes of units with the same
  degree $k$ and with occupation number
  $N_k$. Obviously, $\sum_{k=k_{\mathrm{min}}}^{k_{\mathrm{max}}}N_k=N$
  has to be satisfied. The denominator can be rewritten
\begin{equation}
  \sum_{j=1}^{N}k_j=\sum_{k=k_{\mathrm{min}}}^{k_{\mathrm{max}}}N_k k.  
  \label{eq:rewrite}
\end{equation}
In the limit of large number of nodes ${N\to \infty}$ this expression
becomes $ N \langle k \rangle$ where the symbol $\langle \cdot
\rangle = \sum_k \;\cdot\;p(k)$ assigns averaging over the degree
distribution. 

Hence, the signal function of the $i$th node becomes
\begin{equation}
  f_i(t)=\frac{k_i}{N}r(t). 
  \label{eq:signal}
\end{equation}
Therein, the mean-field amplitude $r(t)$ is defined by
\begin{align}
  r(t) = \frac{1}{\langle k \rangle} \,
  \sum_{k=k_{\mathrm{min}}}^{k_{\mathrm{max}}}\, \frac{N_k}{N}\, k\,
  \frac{1}{N_k} \sum_{j=1}^{N} \, \delta_{k_, k_j} \, {s}_{k_j}(t).
  \label{eq:mfA2}
\end{align}

It is assumed that, after forgetting initial conditions, units
with the same degree share stochastic pulse sequences which are
statistically identical.  Therefore, the same pulse sequence can be
assigned to units of the same degree class by taking the average over
the corresponding class
\begin{equation}
  \overline{s}_k(t)=\frac{1}{N_k}\sum_{j=1}^N\, \delta_{k,k_j}\, s_{k_j}(t).
  \label{eq:sk}
\end{equation}
Therein, the sum runs over $N_k$ items due to the action of the
$\delta$-function. Since the number of nodes with degree $k$ scales as
$N_k \propto N$, application of the limit of large $N$ yields
\begin{align}
  \lim_{N \to \infty} \overline{s}_k(t) = P(1,t|k), 
  \label{eq:thlimes1}
\end{align}
which was introduced in \eqref{eq:Bayes}.

%
% Using \eqref{eq:newoutput}-\eqref{eq:sk} the following identical
% reformulations are made. First, the sum over the edges is split into a
% sum over the different classes of degrees:
% %
% \begin{align}
% \label{eq:mfA1}
%    f_i(t) &= \frac{k_i}{N\sum_{l=1}^{N}k_{l}}\sum_{j=1}^{N}k_j s_j(t) \\
%   \nonumber &=  \frac{k_i}{\sum_{l=1}^{N}k_{l}}
%   \sum_{k=k_{\mathrm{min}}}^{k_{\mathrm{max}}} \frac{1}{N}\sum_{j\in
%     k\mathrm{-class}} k_j s_j(t).
% \end{align}
% %
% In a second step, the averaged input is inserted into all nodes with
% degree $k$
% %
% \begin{align}
% \label{eq:mfA}
% \nonumber f_i(t) &= \frac{k_i}{\sum_{l=1}^{N}k_{l}}
%   \sum_{k=k_{\mathrm{min}}}^{k_{\mathrm{max}}} \frac{N_k}{N}
%   \frac{k}{N_k} \sum_{j\in k\mathrm{-class}} s_j(t) \\
%    &=
%   \frac{k_i}{\sum_{l=1}^{N}k_{l}}\sum_{k=k_{\mathrm{min}}}^{k_{\mathrm{max}}}\frac{N_k}{N}k\,
%   \overline{s}_k(t).
% \end{align}
% %
% Note, that this expression constitutes a weighted mean-field
% approximation as in \cite{Sonnenschein2012}, namely a signal function via
% %
% \begin{equation}
%   f_i(t)=\frac{k_i}{N}r(t),
%   \label{eq:signal}
% \end{equation}
% %
% where the mean-field amplitude $r(t)$ is defined by \eqref{eq:mfA} or
% explicitly as
% %
% %

% For large number of nodes
% %
% \begin{equation}
%   \lim_{N\to\infty}\frac{N_k}{N} = p(k)
%   \label{eq:thlimes}
% \end{equation}
% %
% $p(k)$ is the given degree distribution of the initial network which
% is identical with the distribution in the replaced network. The
% ensemble is precisely given by the corresponding degree class; hence,
% conditioning on this class is introduced. 
Thus, the mean-field $r(t)$ introduced in \eqref{eq:mfA2} becomes in
this limit
\begin{equation}
  r(t)=\frac{1}{\langle
    k\rangle}\sum_{k'=k_{\mathrm{min}}}^{k_{\mathrm{max}}}p(k')k'P(1,t|k)
  =\frac{\langle k\,P_{1,k}(t)\rangle}{\langle k\rangle}.
	\label{eq:r}
\end{equation}

Eventually, by inserting \eqref{eq:signal} and \eqref{eq:mfA2}
via \eqref{eq:r} into the mean-field description \eqref{eq:norm} and
\eqref{eq:dtP2} the following expression is obtained
\begin{align}
\label{eq:MF}
&P(0,t|k)=1-P(1,t|k),\\ \nonumber
&\dot{P}(1,t|k)=\gamma\left[\frac{k}{N}r(t)\right]\left(1-P(1,t|k)\right)-
\\ \nonumber
&\int_0^{\infty}\gamma\left[\frac{k}{N}r(t-t')\right]\left(1-P(1,t-t'|k)\right)w_1(t')\mathrm{d}t'.
\end{align}
This is now a set of coupled non-linear \linebreak
integro-differential equations, since the mean-field $r(t)$ depends on
$ P(1.t|k)$ via \eqref{eq:r}.  Though the structure of the equations
looks similar to the previous Master equation \eqref{eq:dtP2} it is
qualitatively different. As the result of the replacement of the
adjacency matrix, the number of equations has reduced drastically
compared to the system \eqref{eq:norm} and \eqref{eq:dtP2}. The index
$k$ in \eqref{eq:MF} is only running over the different possible
degrees in the network $k\in[k_{\mathrm{min}},k_{\mathrm{max}}]$
whereas in \eqref{eq:norm} and \eqref{eq:dtP2} it runs over all nodes
$N$.  In addition, the dependence of the activation rate $\gamma$ on
the degree $k$ appears uniquely in the argument for all nodes as
linear factor in the signal function, cf.  \eqref{eq:signal}. In the
master equation for nodes with the degree $k$ it reads
 \begin{equation}
   \label{eq:coup}
   \gamma[f]=\gamma\left[\frac{k}{N} r(t)\right].
 \end{equation}

\subsection{Stationary behavior of excitable units}
\label{sec:qual}
Depending on the specific structure of $\gamma$ this equation can be
highly non-linear. Here in this manuscript, the activation rate
$\gamma$ is assumed to follow Arrhenius' law
\cite{Rice1954,Hanggi1990}.  The two-state system shall mimic the
behavior of stochastic excitable dynamics \cite{Lindner2004} with
state $0$ being the rest state and state $1$ the excited one,
respectively. Transitions to the excited state $1$ is achieved by
overcoming a threshold with barrier $\Delta U$ under the
influence of noise with intensity $D$. The corresponding Arrhenius'
law of the rate reads
\begin{equation}
  \gamma  =  \gamma_0 \,\exp\left(-\frac{\Delta U}{D}\right),
\end{equation}
with a constant $\gamma_0$ defining the time scale. 

The coupling between units is assumed to be purely excitatory and thus
each coupled unit that is already in the excited state will lower the
potential barrier by an amount proportional to the coupling strength
$\sigma$, which is the same for every unit throughout this paper. The
following ansatz for the potential $\Delta U$ combines the above
information
\begin{equation}
  \Delta U = 1-\sigma\,\frac{k}{N}r(t).
	\label{eq:g_net}
\end{equation}

Setting $p(k)=\delta_{k,N}$ restores the globally connected network
with $r(t)\rightarrow P(1,t|N)$, which has been earlier studied in
detail \cite{Kouvaris2010}.  Alternatively, it is possible to consider
a discrete number of $K$ degrees $k_i$, $i=1, \dots, K$.  The
corresponding degree distribution follows as
\begin{align}
   \label{eq:dd}
      p(k) =  \sum_{i=1}^{K} \, \nu_i\, \delta_{k_i,k}
\end{align}
Insertion of the specific rate and degree distribution \eqref{eq:MF}
yields a set of $K$ nonlinear master equations.  A qualitative
discussion of the possible stationary solutions $P^*(1|k_i)$ which
will be approached as $t\to \infty$ will be outlined.  Since the
equations are nonlinear there might exist a different number of
stationary solutions with different stability.  If distributed by
\eqref{eq:dd} and with the rate function \eqref{eq:g_net} these
stationary states are defined by the set of $K$ coupled nonlinear
algebraic equations
\begin{align}
        P^*(1|k_i)\, &= \, \frac{\tau\gamma_i^*}{1+\tau\gamma_i^*},
	\label{eq:SS_gen}
\end{align}
with the stationary spiking rates of the elements with degree
  $k_i$
\begin{align}
\label{eq:rate_ss}
\gamma_i^* &= \gamma\left[ \frac{k_i}{N}\,r^* \right]
\end{align}
and the stationary mean-field amplitude with \eqref{eq:SS_gen}
inserted and averaged over the discrete distribution \eqref{eq:dd}
\begin{align}
  \label{eq:ma_ss}
  r^* &= \frac{1}{\langle k \rangle}\, \left\langle k \,P^*(1|k)
  \right\rangle.
\end{align}

The maximal number of possible stable solutions of this set of
equations can be estimated to be of the order $O(2^K)$. The behavior
is similar to a spin chain with $K$ elements.  Every population is
stable in the rest state $s_i=0$. If being coupled, every population
with given degree reaches a stationary probability to be in the
excited state.

The precise number depends on the specific degree values, the noise
intensity $D$ and the coupling constant $\sigma$.  Generally for low
coupling, respectively for high noise only a single solution exists,
which is the disordered state of all populations.  Lowering of noise,
respectively increasing coupling, enlarges the number with multiple
stable states, including inhomogeneous cases and the two homogeneous
situations where all populations are in the rest or in the excited
states. In general, it depends on the initial conditions which state
will be populated.

Corresponding to the selected steady-state-solution the spiking rates
differ. The spiking rate from rest to excited is determined by the
stationary states of the population which the element belongs to.  It
is expressed by the rate given in \eqref{eq:rate_ss} where the
specific solution has to be inserted. Such state dependent
  dynamical behavior of neuronal activity was recently discussed for
  phase oscillators in \cite{Kromer2014}.

In this simplified model no Hopf-bifurcation can take place. 
  All interactions have an aligning effect of the elements similar to a
  spin chain. Therefore, the existence of stable
  oscillating, chimaera state, cluster synchronization or chaotic
  solutions can be excluded. 
  But adding delayed feedback of the mean-field, mixtures
  of excitatory and inhibitory acting units of the network or
  systematic shifts in the signal function might be a source for
  more complex situations.

Special initial conditions (for example all units in the excited
  state) cause damped oscillations of the mean-field. 
  Even in the case of a $\delta$-function as waiting time
  density, the assumed exponentially distributed activation times
  scatter the individual spins with large dispersion. 
  Hence the
  coherence of special initial conditions is destroyed after a few
  excitations yielding a stationary mean-field.

Nevertheless, as will be seen in the next section, the
  individual spins can fire with a small CV\footnote{The coefficient of variation (CV) is given by the ratio of the standard deviation to the mean of the distribution.} resembling oscillatory
  behavior. In the states with high mean-field values the activation time
  becomes negligible small. In these situations the spiking is
  dominated by the recovery time from the excited to the rest
  state. If this time does not possess remarkable dispersion the CV
  become vanishingly smalls.

In the next Section \ref{sec:RBN} further insight into
  the consequences of several degrees will be given by dealing with the simplest
  case of network with two populations with different degrees
  $k_1, k_2$, only.

\section{Random binary networks}
\label{sec:RBN}
In the following, the properties of a binary random network
\cite{Lambiotte2007,Sonnenschein2013,Carro2016} will be studied. It
allows a full sketch of the possible bifurcation scenario appearing in
these networks. These ones are randomly connected with two different
degrees $k_1$ and $k_2$.

%This work is focused on the limit $N \to\infty$ in dense networks. 
%Hence, one may introduce the ratio
%%
%\begin{equation}
%  \label{eq:beta}
%  k =  \frac{k}{N}  
%\end{equation}
%%
%which is used instead of $k$ and $N$, since the degrees scale with the
%number of nodes in dense networks~\footnote{As shown in
%  \cite{Sonnenschein2012} such scaling is necessary for the validation
%  of the heterogeneous mean-field approximation used in this section}.
%The value $k$ is known as connectivity fraction or degree density.
%Its values are distributed by the density which we again denote
%$p(k)$, for simplicity. Consequently, averages are now defined as
%$\langle\cdot\rangle = \int_0^\infty p(k)\cdot\mathrm{d} k$.

In terms of the degree distribution for a binary network it is
supposed that
\begin{equation}
	p(k)=\nu\delta_{k,k_1}+(1-\nu)\delta_{k,k_2}.
	\label{eq:pbeta_rbnet} 
\end{equation}
where $\nu\in]0,1[$ is the fraction of nodes with degree $k_1$. In
this paper $k_1 > k_2$ is set, but due to the symmetry
$(k_1, \nu) \longleftrightarrow (k_2, 1-\nu)$ the alternative
case is also included.

The set of two master equations for binary networks read:
\begin{align}
  \nonumber &\dot{P}(1,t | k_1) = \gI(t)\;(1-P(1,t|k_1)-\\
  \nonumber &\int_0^\infty \gI(t-t') (1-P(1, t-t'|k_1))w_1(t') \mathrm{d} t',\\
  \nonumber &\dot{P}(1,t | k_2) = \gII(t)\;(1-P(1,t|k_2))-\\
  &\int_0^\infty \gII(t-t') (1-P(1,t-t'|k_2))w_1(t') \mathrm{d}t',
	\label{eq:MFeq_rbnet}
\end{align}
where we have denoted $\gamma_{k_1}(t)=\gamma[k_1 / N r(t)]$ and
$\gamma_{k_2}(t)$, respectively. Equations \eqref{eq:MFeq_rbnet}
can be \linebreak brought into the integral form \cite{phdKouvaris}:
\begin{align}
	\nonumber P&(1,t|k_1) = \\
	\nonumber &\int_0^\infty \gI(t-t') (1-P(1,t-t'|k_1))z_1(t') \mathrm{d} t',\\
	\nonumber P&(1,t|k_2)  = \\
	 &\int_0^\infty \gII(t-t') (1-P(1, t-t'|k_2))z_1(t') \mathrm{d}t',
	\label{eq:MFeqInt_rbnet}
\end{align}
supplemented by initial conditions. Therein, $z_1(t)$ is the survival
probability of state $1$,
\begin{equation}
	z_1(t) = 1-\int_0^{t}w_1(t') \mathrm{d} t'.
\end{equation}
Equations \eqref{eq:MFeqInt_rbnet} have to be supplemented by initial
conditions.

\subsection{Qualitative discussion of steady states}
\label{subsec:SS}

Equations \eqref{eq:MFeqInt_rbnet} are suitable for calculating the
steady states of this system. For a steady state
$\lim_{t\rightarrow\infty} P(1,t | k_i)=P^*(1 | k_i)$ applies.  Using
\eqref{eq:MFeqInt_rbnet} and integration by parts gives the following
coupled implicit equations for the steady states
\begin{align}
	\PbetI &= \frac{\tau\gI^*}{1+\tau\gI^*},\qquad\PbetII = \frac{\tau\gII^*}{1+\tau\gII^*}.
	\label{eq:SS_rbnet}
\end{align}
The values of $\PbetI^*$ and $\PbetII^*$ define the stationary order
in the two subpopulations. In Eq. \eqref{eq:SS_rbnet} we introduced
the mean relaxation time of the excited state $\tau=\int_0^\infty
t\,w_1(t)\mathrm{d} t$ and the steady state activation rate
$\gamma^*_{k_i}=\gamma[k_i r^*]$ for $i=1,2$ depending on the steady
state mean-field $r^*$. Equation \eqref{eq:r} defines the order
parameter of the full network. It becomes $r^*=\left\langle k P^*(1 |
  k)\right\rangle / \langle k \rangle $. Taken at steady state, this
yields a transcendent equation for the steady state value $r^*$ of the
mean-field which yields
\begin{equation}
r^*= \frac{1}{\langle k \rangle}\left\langle \frac{k}{1+\frac{1}{\tau\gamma_{k}^*}}\right\rangle.
\label{eq:rSS_rbnet}
\end{equation}
This equation can possess several solutions which we will discuss in
detail, later on. In case of large noise $D \to \infty$, only the
homogeneous disordered solution $r^*=1/2$ exists.  In this limit the
exponential function becomes unity and since we will select $\gamma_0
\tau \approx 1$ the disordered state is characterized by $r^*=1/2$.  A
constant value of $r^*$ does not imply that the activity of the
individual nodes has ceased.  It is rather the mean activity or flow
that is constant (cf.~Fig.~\ref{fig:xAS}).  Oscillating behavior of
$r^*$ would correspond to synchronization among the units.  But
without further ingredients like delayed feedback or additional
inhibitory coupled nodes such states cannot be reached.

The transition to the disordered state can be studied in more
detail. Demanding that the first derivatives of l.h.s. and r.h.s.
with respect to $r$ coincide at $r=r^*$ , provides a condition for a
saddle-node bifurcation.  The homogeneous disordered state becomes
unstable and two new stable solutions occur.

Execution of the derivatives in \eqref{eq:rSS_rbnet} results in
\begin{align}
	\frac{k_1
		\nu}{\langle k\rangle}\frac{\tau\;\frac{\partial\gI^*}{\partial 
			r^*}}{(1+\gI^*\tau)^2} + \frac{k_2
		(1-\nu)}{\langle k\rangle}\frac{\tau\;\frac{\partial\gII^*}{\partial 
			r^*}}{(1+\gII^*\tau)^2} &= 1.
	\label{eq:SNEllipseinR_rbnet} 
\end{align}
%
% where the shorthand notation\linebreak
% $\partial\gamma^*_{k_i}/\partial r^*=
% \partial\gamma_{k_i}/\partial
% P(1 | k_i)\big|_{P(1 | k_i)=P^*(1 | k_i)}$ is used. 
Using \eqref{eq:SS_rbnet} this becomes
\begin{align}
	\nonumber 1=\frac{k_1 \nu}{\langle k\rangle}(\PbetI-(\PbetI)^2)\frac{\partial \log(\gI^*)}{\partial r^*} &+\\
	\frac{k_2 (1-\nu)}{\langle k\rangle}(\PbetII-(\PbetII)^2)\frac{\partial \log(\gII^*)}{\partial r^*} &.\\
\end{align}
Changing the variables to $x_{k_1}:=\PbetI-\frac{1}{2}$ and
$x_{k_2}:=\PbetII-\frac{1}{2}$ and rearranging the equation
gives
\begin{align}
	\frac{x_{k_1}^2}{a_1^2} +\frac{x_{k_2}^2}{a_2^2} &= 1.
	\label{eq:SNEllipse_rbnet}
\end{align}
Equation \eqref{eq:SNEllipse_rbnet} defines an ellipse with semi-axes
\begin{equation}
		a_1 = \frac{1}{2}\sqrt{\frac{ \big\langle k\frac{\partial
					\log(\gamma^*_{k})}{\partial
					r^*}\big\rangle-4\langle k\rangle}{k_1 \nu \frac{\partial
					\log(\gamma_{k_1}^*)}{\partial r^*}}}
\end{equation}
and similarly $a_2$ with the substitutions\linebreak $k_1
\rightarrow k_2$ and $\nu \rightarrow (1-\nu)$.  The ellipse
reduces to a point where the two bifurcations merge.  It corresponds
to $\PbetI~=~\PbetII~=~r^*~=~\frac{1}{2}$, at
\begin{equation}
\left\langle k\frac{\partial \log(\gamma^*_{k})}{\partial r^*}\right\rangle-4\langle k\rangle = 0.                                 
\label{eq:SNEllipseCrit_rbnet}
\end{equation}
For the $\gamma_k$ given by \eqref{eq:g_net},
\eqref{eq:SNEllipseCrit_rbnet} results in
\begin{equation}
\frac{\sigma_{\text{crit}}}{4 D_\text{crit}}\frac{\langle k^2\rangle}{N\,\langle k\rangle}=1.
\label{eq:scusp_rbnet}
\end{equation}

Comparing this to the well known result of all-to-all coupled
stochastic two-state units \cite{Kouvaris2010},
\begin{equation}
\frac{\sigma_{\text{crit}}}{4 D_\text{crit}}=1,
\end{equation}
it is visible that it differs only by a scaling factor given by the
ratio of the first two moments of the degree (density) distribution.
This is a typical network effect in mean-field coupled oscillators
\cite{Restrepo2005,ArenasKuths08,Boccaletti06,Newman03}.  The factor
can be interpreted as the mean of the degree distribution of the
nearest neighbors \cite{Dorogovtsev02} assuming that the local
structure of the network is treelike. Therefore the effective coupling
strength
\begin{equation}
\sigma_{\rm eff} = \sigma\frac{\langle k^2\rangle}{\langle k\rangle} 
= \sigma\frac{\nu k_1^2 + (1-\nu) k_2^2}{\nu k_1 + (1-\nu) k_2} 
\label{eq:sstar_rbnet}
\end{equation}
will be introduced. Note that it is evident from
\eqref{eq:SNEllipseCrit_rbnet} that this scaling is only obtained if
$\gamma$ depends exponentially on the mean-field $r$.

\begin{figure*}[htb!]
	\centering
        \includegraphics[width=.48\textwidth]{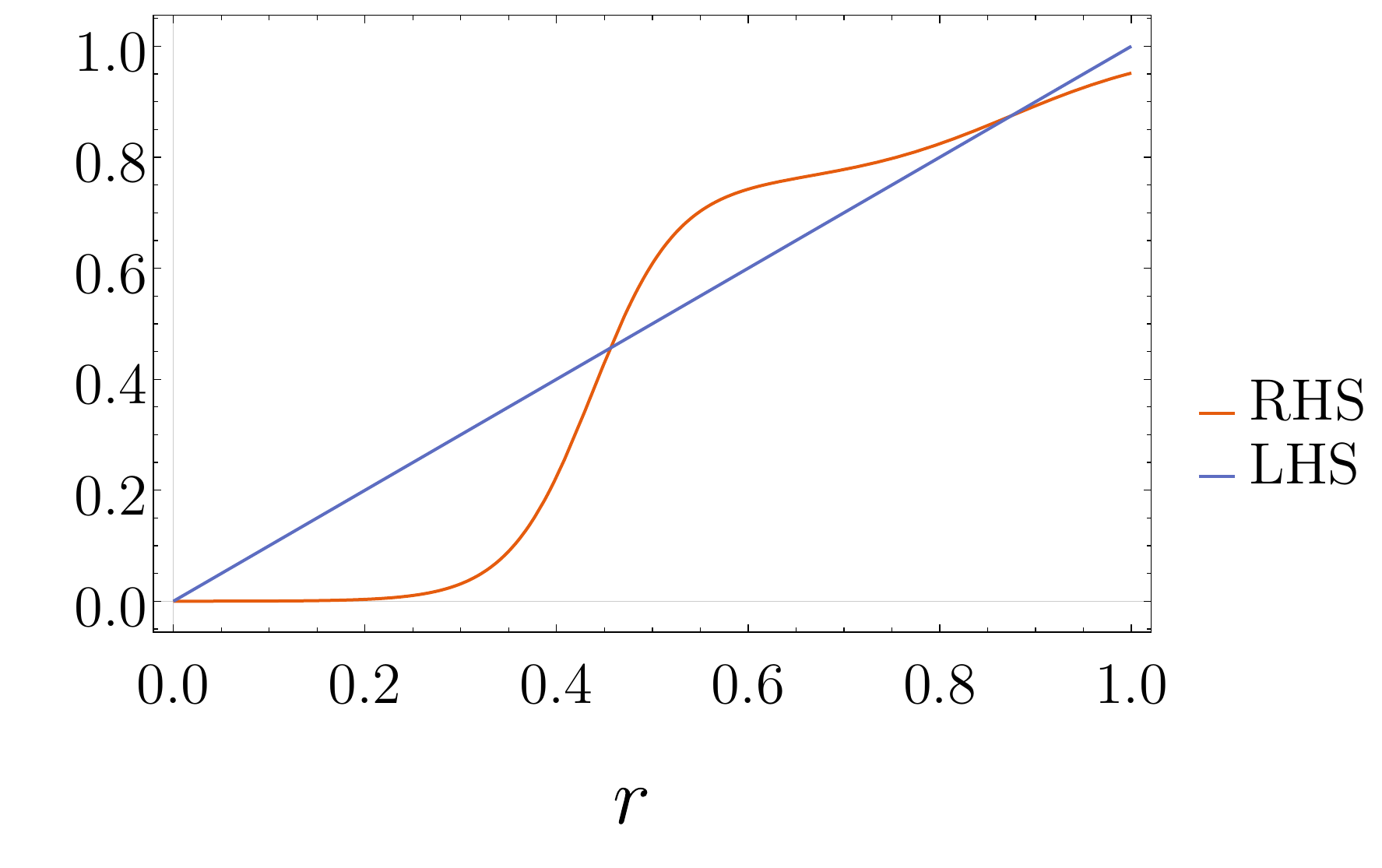}\includegraphics[width=.48\textwidth]{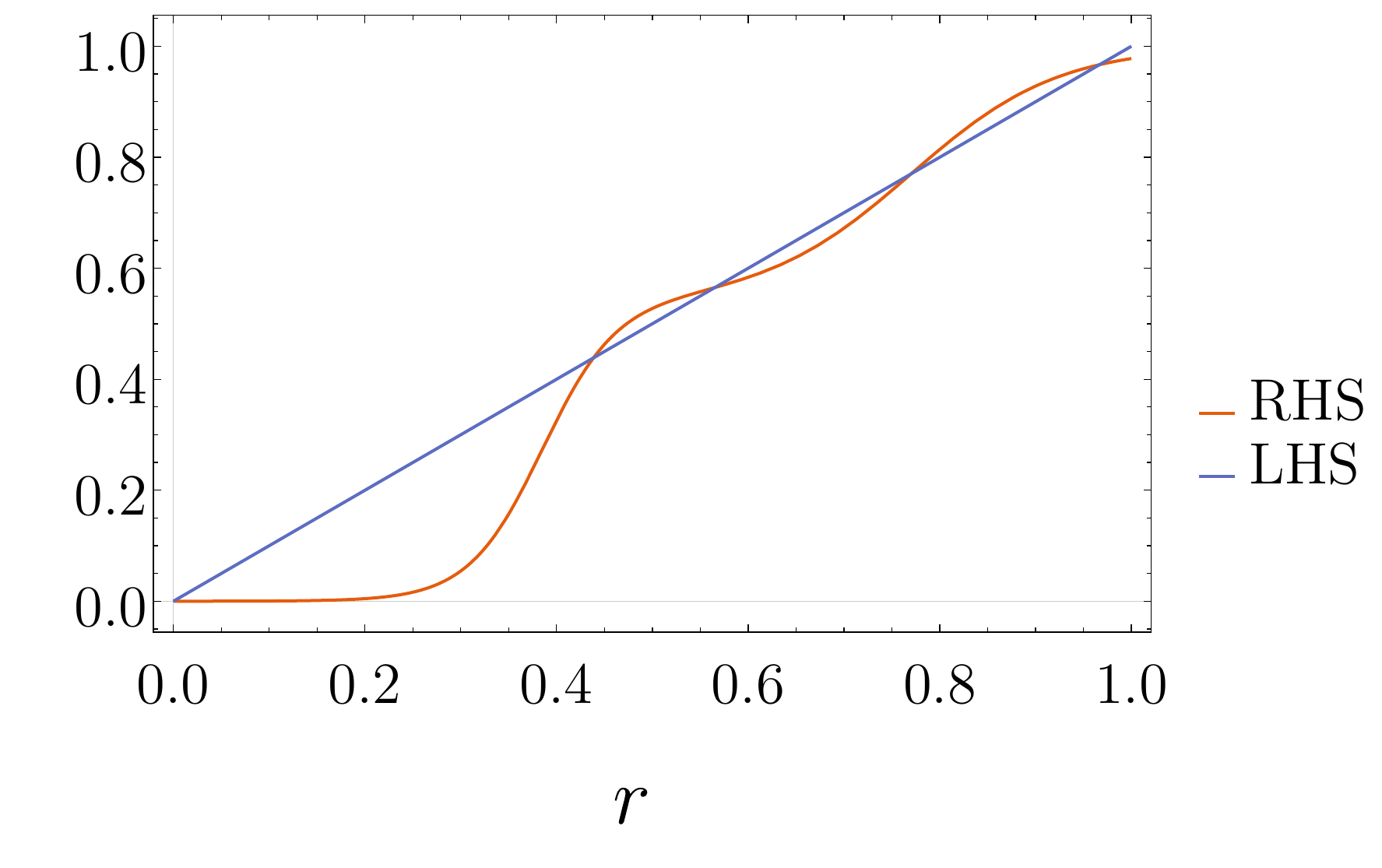}
    \caption{Graphical representation of left- and right-hand side
      (LHS, RHS) of \eqref{eq:rSS_rbnet}. $D=0.1$, $k_1 / N = 1/2$,
      $k_2 / N=1/4$, $\gamma_0\cdot\tau=1$. Left: $\nu=0.6$,
      $\sigma\approx4.57$, leads to three fix-points; two are stable
      and one is unstable. Right: $\nu=0.37$, $\sigma\approx5.19$
      gives rise to five fix-points; three stable ones and two
      unstable ones.}
	\label{fig:rvsl}
\end{figure*}

In case of low noise a more detailed picture with possibly multiple
solutions and ordered states occur. These solutions can be discussed
solving \eqref{eq:rSS_rbnet} graphically and plotting the
r.h.s. versus the l.h.s as presented in Fig.~\ref{fig:rvsl} for
typical situations.

The l.h.s. of equation \eqref{eq:rSS_rbnet} is a straight line and
unbounded whereas the r.h.s. grows monotonically and is bounded
between values of the interval $[0,1]$. Hence solutions $r^*$ are also
in this interval. Solutions with one, three or five intersections can
be found. Bifurcations between these monostable, bistable of tristable
behavior are saddle-node bifurcations or, if these coincide, a
pitchfork bifurcation.

If the number of degrees would be increased, the number of steps will
also increase in the same manner, giving rise to even higher
multistable states.

\subsection{Solutions with vanishing noise}
\label{subsec:D0}
It is illustrative to look first in detail at the case of vanishing
noise.  Then the r.h.s. of equation \eqref{eq:rSS_rbnet} vanishes
$\propto \exp(-1/D)$ as $r \to 0$ and approaches unity for large
values of $r$. In between, the r.h.s. makes two jumps with magnitude
$\nu k_1/\langle k \rangle $ and $(1-\nu)k_2/\langle k
\rangle$, respectively. These steps are located at $r_1=1/(\sigma
k_1)$ and $r_2=1/(\sigma k_2)$.  For the r.h.s. to possess one
intersection (monostability) with the straight line $r$, we obtain the
following conditions by using that $k_1 >k_2$:
\begin{equation}
\label{eq:twosteps}
\frac{1}{\sigma k_1}\, > \, \frac{\nu k_1}{\nu k_1+(1-\nu)
	k_2},\,\qquad\frac{1}{\sigma k_2} >  1.
\end{equation}
For vanishing noise the monostable state is always the ordered one
with $r^*=0$, i.e. neither of the two populations is excited..  If one
of these inequalities is violated, the mean-field dynamics exhibit
bistability. Given that the first one does not hold, besides the
ordered solution with $r^*=0$, a second stable inhomogeneous state
appears. The higher degree population is in the excited state and the
lower degree population remains in the non-excited one
(cf. Fig.~\ref{fig:1k2_1k8}).  In contrast, if the second inequality
is violated bistability occurs between the homogeneous non-excited
states and the homogeneous excited ones (cf. Fig.~\ref{fig:1k2_1k3}).
Finally, if both inequalities do not hold, the solution has five
intersections according to a tristable solution between the two
homogeneous situations and the one non-homogeneous one
(cf. Fig.~\ref{fig:1k2_1k4}).

\subsection{Solutions with finite noise and simulations}
\label{subsec:sim}
\begin{figure*}[htb!]
  \centering
  \begin{subfigure}{\linewidth}
  	\includegraphics[width=1\textwidth]{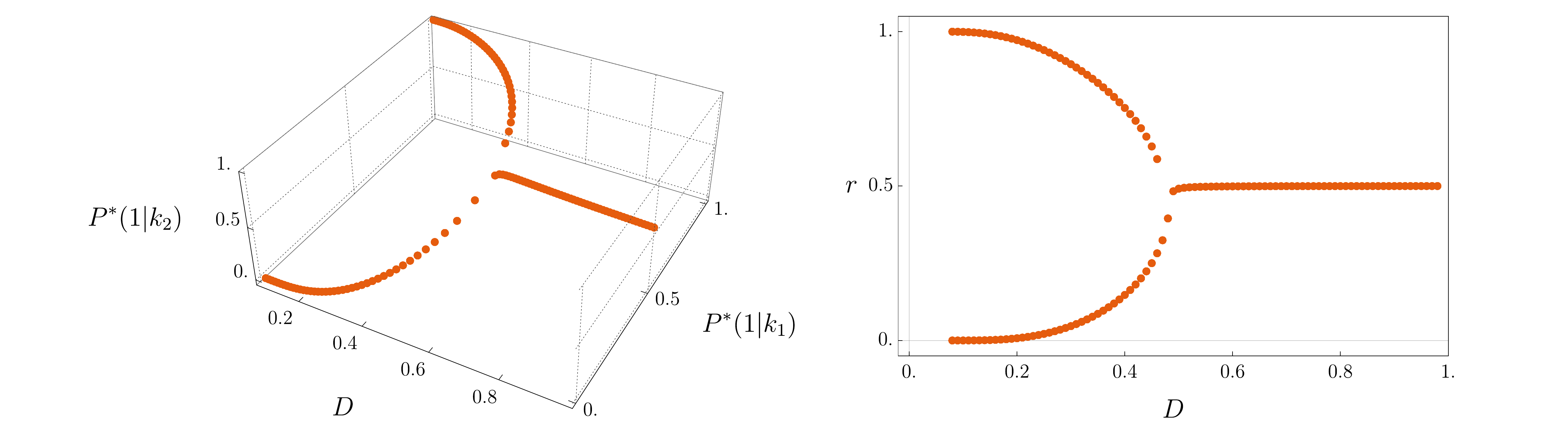}
  	\caption{}
  	\label{fig:1k2_1k3}
  \end{subfigure}
  
  \begin{subfigure}{\linewidth}
  	\includegraphics[width=1\textwidth]{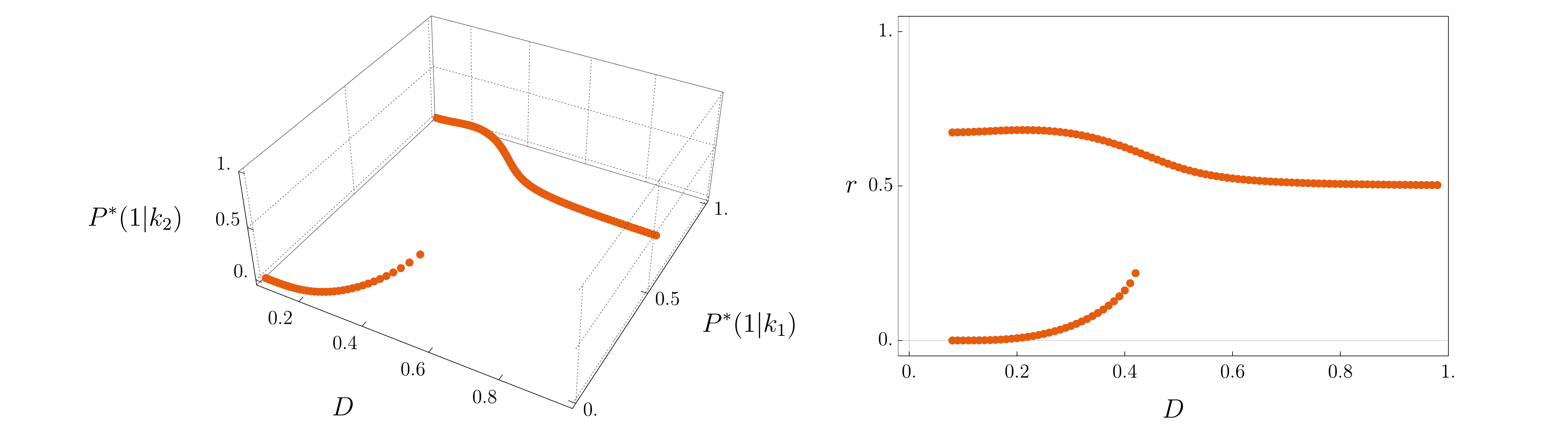}
  	\caption{}
  	\label{fig:1k2_1k8}
  \end{subfigure}
  
  \begin{subfigure}{\linewidth}
  	\includegraphics[width=1\textwidth]{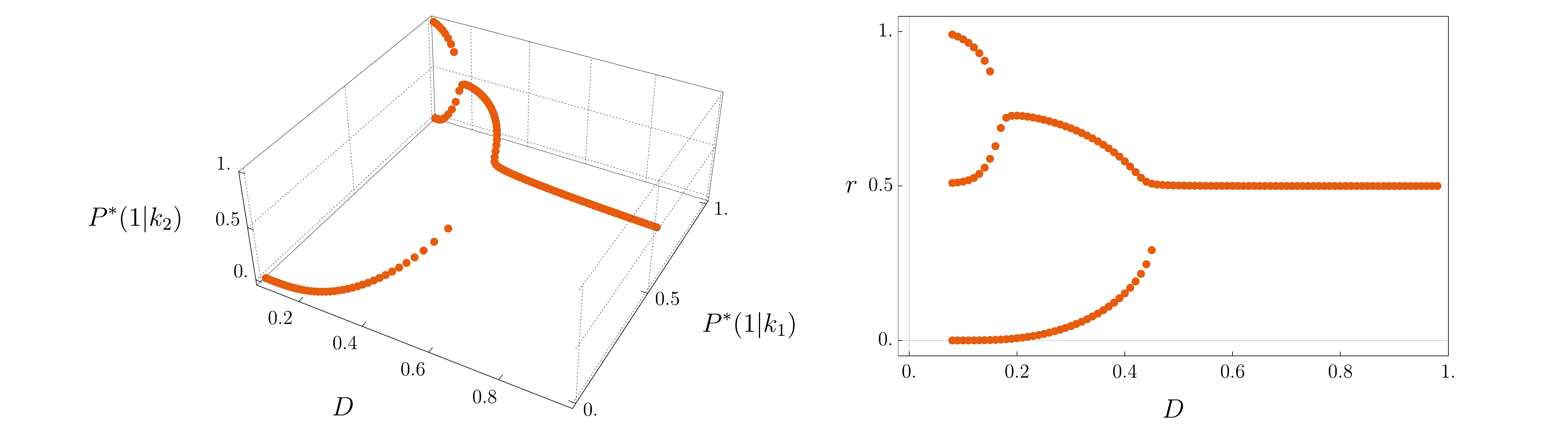}
  	\caption{}
  	\label{fig:1k2_1k4}
  \end{subfigure}
  \caption{Stable branches of \eqref{eq:MFeq_rbnet} for $\sigma_{\rm eff}=2$,
    $\gamma_0\cdot\tau=1$, $\nu=0.34$, $k_1 / N=\frac{1}{2}$ and
    varying $k_2 / N$.
    \newline\subref{fig:1k2_1k3}) $k_2 / N=\frac{1}{3}$: 
    The typical bifurcation into two homogeneous states. 
    This is also observed in networks with all-to-all topology.
    \newline\subref{fig:1k2_1k8}) $k_2 / N=\frac{1}{8}$:
    This bifurcation is similar to
    Fig.~\ref{fig:1k2_1k3} but the lower degree population cannot be
    in the excited state anymore.
    \newline\subref{fig:1k2_1k4}) $k_2 / N=\frac{1}{4}$:
    In this bifurcation diagram
    there is a monostable regime for large $D$, a bistable regime
    with a total ordered and a partial ordered state for intermediate
    $D$ and the tristable state with three different ordered states
    for low $D$.
    }
  \label{fig:1k2_1kx}
\end{figure*}

Examples of the qualitative behavior of the stea\-dy states for
various noise levels $D$ are presented in
Figs.~\ref{fig:1k2_1k3}-\ref{fig:1k2_1k4}. The graphs have been
obtained by numerical solution of \eqref{eq:rSS_rbnet} and
\eqref{eq:SS_rbnet}. The main difference of these graphs is the ratio
$\alpha=k_1 / k_2$. In Fig.~\ref{fig:1k2_1k3}, $\alpha$ equals
$3 / 2$ and for high noise the disordered state with mean-field $r^*=1
/ 2$ is stable. The latter bifurcates for lower noise to the known
bistability of ordered states \cite{Kouvaris2010} which approach
$(0,0)$, $(1,1)$ as $D \to 0$. These have been written in terms of the
state vector $\Pvec(t)=(P(1,t | k_1),P(1,t |  k_2))^\mathrm{T}$. With
respect to the network these solutions are homogeneous states since
both populations of the network are ordered in the same states.

Figure~\ref{fig:1k2_1k8} presents the qualitative behavior with a
strong mismatch of the degrees of the two populations, namely
$\alpha=4$.  The graph shows that in this parameter region bistability
of the two ordered states occurs for lower noise values. With
vanishing $D$-values the states become $(0,0)$ and $(0,1)$.  In
difference to the previous case, the second solution $(0,1)$ is
inhomogeneous with respect to the two populations in the network. In
the $(0,1)$ state one population is ordered in state $0$ whereas the
other approaches an ordered state with mean activity $1$. It is a
result of the strong mismatch $\alpha$ of the degrees and of the
asymmetry $\nu$ of the two populations. If, for example, the first
smaller population with a higher degree is ordered in the excited
state $1$, it is not able to excite the second larger population
anymore. The latter remains in the ordered rest state $0$.

Also the coexistence of both scenarios is possible and give rise to a
tristable parameter region as presented in Fig.~\ref{fig:1k2_1k4}.
Here a moderate value of $\alpha=2$ was selected. Lowering the noise
intensity, three different regions are visible. First, for high noise
$D > 0.5$ the mono\-stable disordered solution exists which is
apparent in all figures. Lowering the noise this state becomes
bistable between the ordered homogeneous states where both populations
are in state $0$ and an inhomogeneous network where the population
that is smaller and stronger connected is with high probability in the
ordered state $1$ but the larger less connected population is still
disordered. This is another type of bistability, this time between an
homogeneous ordered state and an inhomogeneous disordered state.
Finally, by decreasing the noise intensity $D$ further, the third
region is entered and the solutions become tristable between $(0,0)$,
$(1,0)$ and $(1,1)$ in case of vanishing noise.  Stability of the
state $(1,0.5)$ as visible in the intermediate region of
Fig.~\ref{fig:1k2_1k4} is an interesting event, because it means that
the population of the network with the higher degree is in an ordered
state whereas the population with the lower degree is in a disordered
state. Such partly ordered states have also been reported for the
Ising model on correlated scale-free networks \cite{Zhou2007}.  These
should not be confused with a chimera state, because the units of the
subpopulations are not identical and the value of $r$ does not reflect
the synchrony of the phases among the units.

\begin{figure*}[htb!]
	\centering
	\includegraphics[width=1\textwidth]{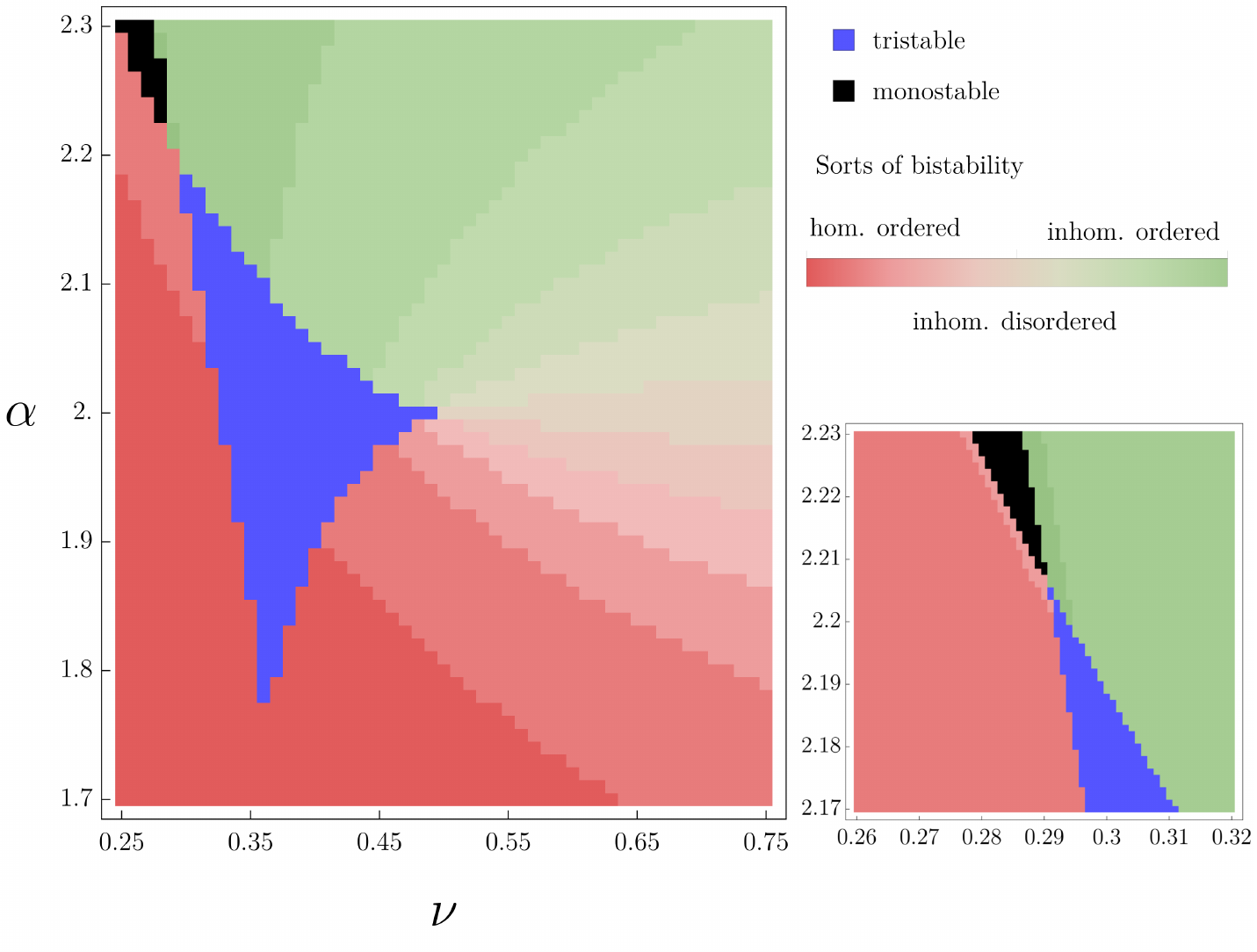}
	\caption{Qualitative behavior of the dynamical regimes in the
          $\alpha-\nu$ plane for $D=0.1$, $\sigma_{\rm eff} =2$ and
          $k_1 / N=1/2$.  The graphic beneath the legend is an excerpt
          of the region between tri- and monostability The two regions
          will meet in a point which is below resolution.}
	\label{fig:alpha_nu}
\end{figure*}

In Fig.~\ref{fig:alpha_nu} the distribution of the different stability
regimes for varying degree mismatch $\alpha$ and relative
concentration $\nu$ are shown for $D=0.1$. The tristable region forms
an island surrounded by the different types of bistability and
connected to the monostable shore by a very narrow region. Going
around the island in an anti-clockwise manner one starts at the
homogeneous ordered configuration which gradually becomes more
disordered and inhomogeneous with maximal disorder in the middle of
the right hand side. After the turning point it gets ordered again but
this time in the inhomogeneous regime.
%
%\FloatBarrier

The presented findings have been confirmed by microscopic simulations
of the coupled network, see Eqs. \eqref{eq:flows} with $w_0(t)$ from
\eqref{eq:w0}.  Numerical investigation of the random binary network
is done by solving the Master equations \eqref{eq:MFeq_rbnet} in the
Markovian case, namely $w_1(t)$ as in \eqref{eq:w1}. As shown in
\eqref{eq:SNEllipseinR_rbnet} and \eqref{eq:SS_rbnet}, the equations
depend on the first moment of the waiting time distribution rather
than the shape of the distribution, although higher moments may play a
role for other bifurcations (cf.  Appendix~\ref{sec:app}).  In
addition microscopic simulations of a random binary network with 6000
nodes and full network topology, which corresponds to \eqref{eq:dtP2},
confirm the made approximations.

\begin{figure*}[htb!]
  \centering
  \includegraphics[width=1.\textwidth]{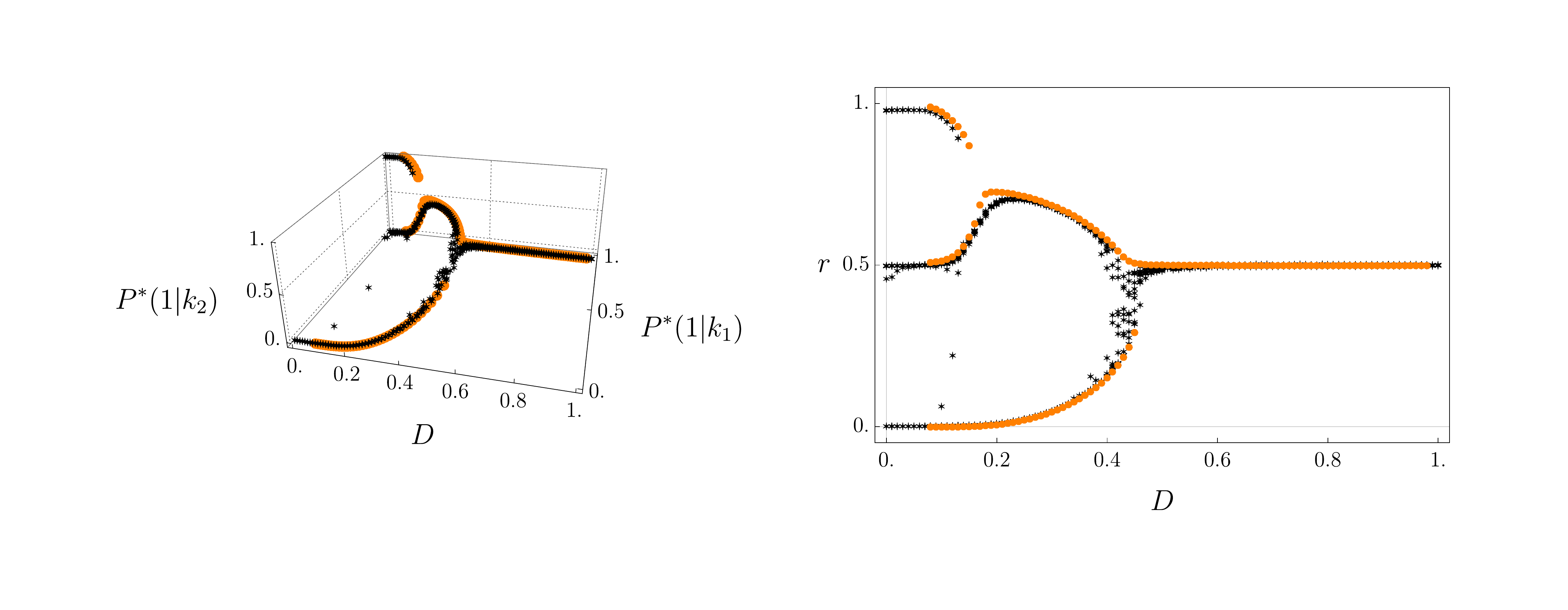}
  \caption{Black stars show the most probable states of the network in
    a microscopic simulation with $k_1=500$, $k_2=250$, $N=6000$,
    $\sigma_{\rm eff}=2$, $\gamma_0\cdot\tau=1$, $\nu=0.34$ and
    $w_1(t)$ from \eqref{eq:w1} after $20\,000$ simulation steps.  The
    orange dots result from the numerical solution of
    Eq.~\eqref{eq:MFeq_rbnet}.  There are small deviations from
    Fig.~\ref{fig:1k2_1k4} which are near the critical points where
    finite time effects are the strongest. For each $D$-value $11$
    equally distributed starting conditions were chosen by preparing
    the network such that $r \in [0,1]$.  }
  \label{fig:MC_1k500_2k250}
\end{figure*}

\begin{figure*}[htb!]
	\centering
        \includegraphics[width=.5\textwidth]{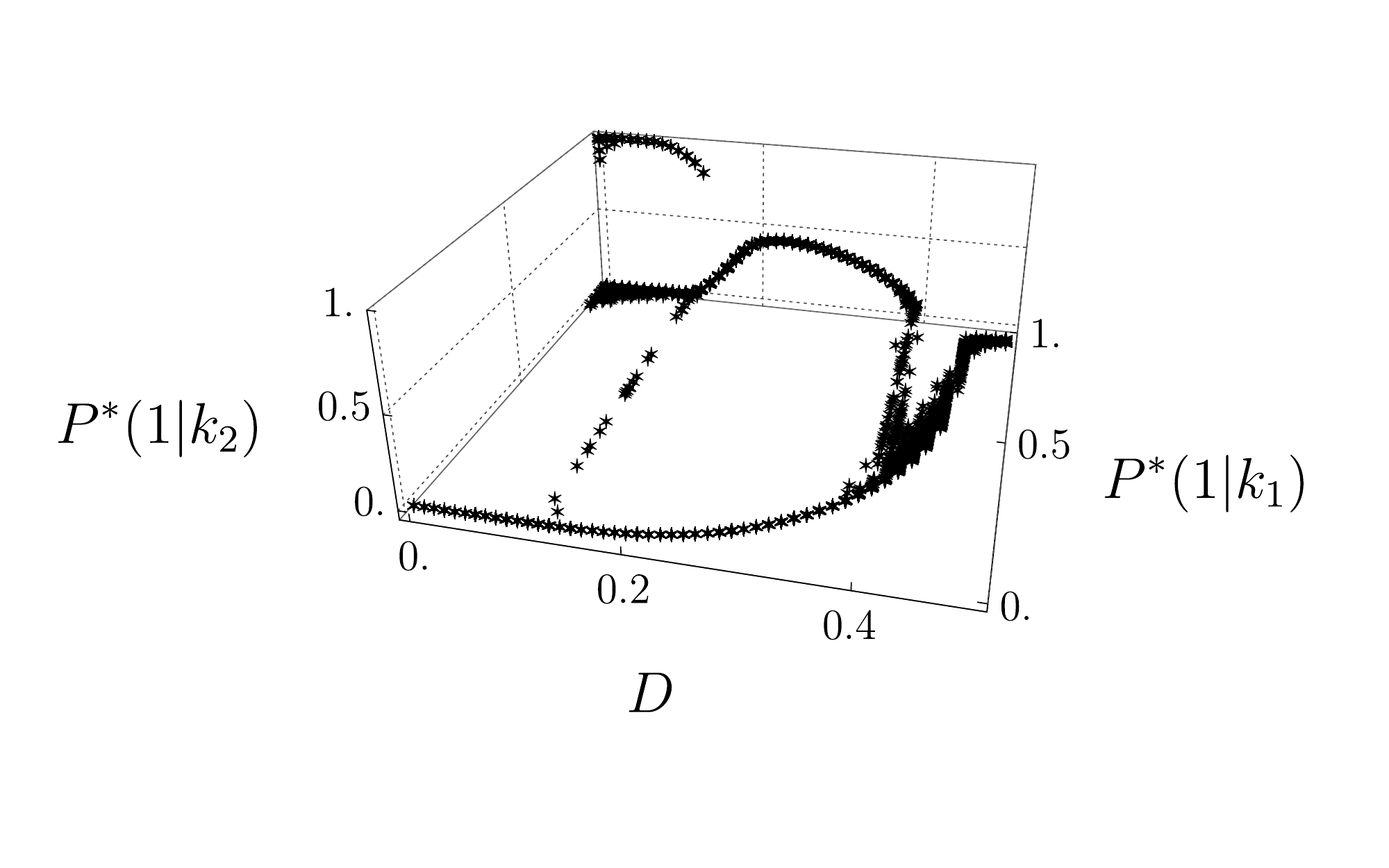}
  \caption{Most probable states of the network in a microscopic
    simulation in the region $D\in[0,0.5]$ with $w_1(t)$ from
    \eqref{eq:w1peak} and other parameters as in
    Fig.~\ref{fig:MC_1k500_2k250}.  But in this figure for each
    $D$-value $121$ equally distributed starting conditions were
    chosen by setting $P(1, 0 |k_1) \in [ 0, 1 ]$ and $P(1, 0 |k_2)
    \in [ 0, 1 ]$ independently.  It shows that the existence of tri-
    and bistability does not depend on the specific choice of
    $w_1(t)$.  }
  \label{fig:simDelta}
\end{figure*}

Exemplary results are shown in Fig.~\ref{fig:MC_1k500_2k250} -
\ref{fig:simDelta} which reproduces Fig.~\ref{fig:1k2_1k4} with two
different waiting time distributions. The exponential waiting time
distribution which was also used in Fig.~\ref{fig:1k2_1k4} and a
$\delta$-distribution with same mean but no variance.

\begin{figure*}[htb!]
	\centering
	
	\begin{subfigure}{0.49\linewidth}
		\includegraphics[width=\textwidth]{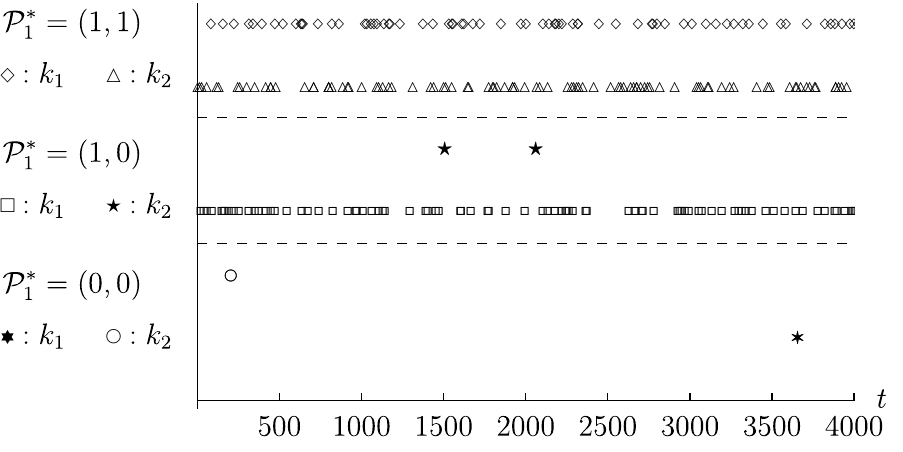}
		\caption{}
		\label{fig:EAS}
	\end{subfigure}
	\begin{subfigure}{0.49\linewidth}
		\includegraphics[width=\textwidth]{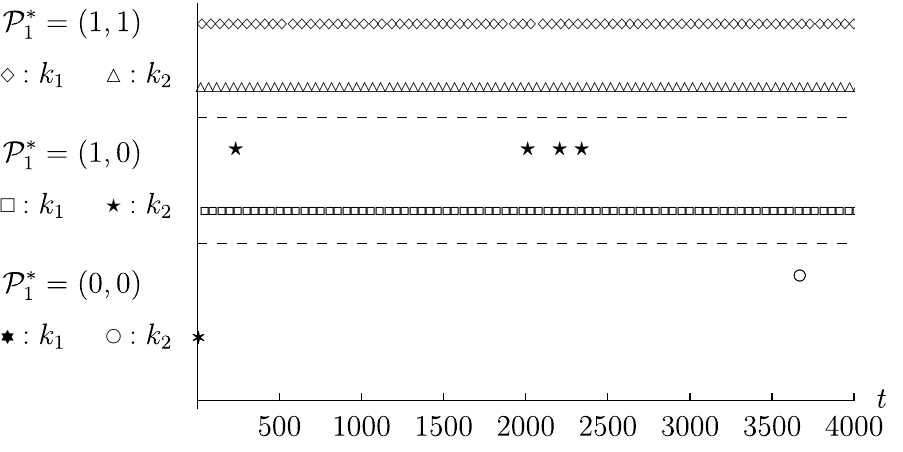}
		\caption{}
		\label{fig:DAS}
	\end{subfigure}
	\caption{
		Activity of a randomly picked node of each degree of the
		networks simulated in Figs.~\ref{fig:MC_1k500_2k250} -
		\ref{fig:simDelta} at $D=0.1$ in the three different steady
		states. 
		Homogeneously activated $(1,1)$, homogeneously at rest $(0,0)$ and inhomogeneously ordered $(1,0)$.
		\newline \subref{fig:EAS}) Network with exponential $w_1$
		corresponding to Fig.~\ref{fig:MC_1k500_2k250}.
		\newline \subref{fig:DAS})
		Network with $\delta$-distributed $w_1$ corresponding to
		Fig.~\ref{fig:simDelta}.
	}
	\label{fig:xAS}
\end{figure*}

Fig.~\ref{fig:xAS} shows the activity of arbitrary nodes in the
  two populations in the three different regimes of stable
  states. The spiking activity of the nodes is presented as symbols
  versus running time accordingly to the noted degree values $k_1,k_2$
  at the r.h.s of the graph. 
  %In addition, also the steady states of the two populations are listed there. 
%  In the upper and lower graphs
%  both populations are homogeneously in the excited or, respectively,
%  in the resting states. The situation in between corresponds to the
%  inhomogeneous steady state where the first population is in the
%  excited state whereas the second population is resting state. 
%

The firing activity in the two different states is different. As
discussed already earlier in Section \ref{sec:qual}, units fire seldom
in the rest state, but rapidly in the excited state. This dynamical
behavior survives also in the inhomogeneous case. In
Fig.~\ref{fig:EAS} we present the activity of the units, with an
exponential relaxation waiting time distribution.  High disorder of
the spiking activity in the excited states is the consequence.  The CV
of the simulated activity is close to $1$, which is the value for an
independent Poissonian spike train. Differently, in Fig.~\ref{fig:DAS}
spiking events from simulations with a $\delta$-distribution are
shown. For states where a population $i=1,2$ is in the excited states,
i.e. if $P(1|k_i) \approx 1$, the measured CV possesses values close
to zero . This corresponds to a perfectly oscillatory behavior of the
units. The period of this spiking coincides with the time $\tau$ the
unit stays in the excited state. After this period the unit flips to
the rest state. The exponentially distributed time to flip back in the
excited state vanishes and also its variance. In consequence, the
units behave loike oscillators. 

It is important to stress that the constant mean-field value does not
correspond to synchronization of the individual units in the excited
states, even in case when they practically oscillate. In the rest
state the measured CV for both choices of waiting time densities are
close to $1$.

\section{Conclusion}
\label{sec:summ}

In this paper, we have investigated\linebreak semi-Markovian
stochastic two-state units embedded in a complex network.  A
theoretical framework has been developed through a heterogeneous
mean-field approximation, which is valid for random uncorrelated
networks. Our work thus represents an extension of previous studies on
globally coupled two-state systems (especially \cite{Kouvaris2010},
but also \cite{Pinto2014,EscaffLind12,Dumont2014}) to two-state
systems that have a complex coupling structure.

As an example, we have focused on a random binary network. Thereby we
have discovered qualitative changes in the behavior of the steady
states.  Specifically, structurally new conformations have been found,
including tristable and partially ordered states. Additionally the
influence of the network on the critical coupling strength has been
revealed. We have corroborated all our theoretical results via
numerical simulations. We found that the ``Markovianity'' of the
underlying process has no great effect on the positions and the number
of steady states and their bifurcations in this simple setting, but
still their basins of attraction may be different. Instead the first
moment of the waiting time distribution has the greatest impact and
higher moments occur only in bifurcations that have at least
co-dimension two.  It remains for future studies to pursue our
analysis explicitly in cases, where more than two different degrees
exist. It will be particularly interesting to see how our findings
regarding the multistability will generalize.

Networks of stochastic two-state units can be seen as a toy model for
magnetic spins, neurons, blinking phenomena or two valued opinions.
The occurrence of tristability is also reported in molecular switches
\cite{Feng2013,Lu2013} and in systems of polaritons
\cite{Cerna2013,Paraiso2010} giving hope to perform ternary logical
operations in the future.  Hence, we expect that our results are
relevant for these real-world systems, where the model considered here
may serve as an idealized version. The analytic tractability is a
strength of our system, but we believe that still many important
extensions await consideration, e.g. including network correlations,
more sophisticated waiting time distributions or coupling functions.

As underlined previously, the main assumption behind the
  heterogeneous mean-field approximation is the lack of degree
  correlations. Random binary networks tend to be
  disassortative, i.e. nodes with different degrees are preferentially
  connected as discussed in \cite{Sonnenschein2013}. However, for the
  network examples considered here and for the chosen parameters,
  these degree correlations become negligible. Therefore, the key
  assumption is not violated in our study which gave the reason for
  our analysis. Extending our theory towards correlated networks
  remains a challenging open problem.  

\begin{acknowledgement}
	The authors thank the TSD/TSP working groups of the Department
        of Physics of the Humboldt University at Berlin for their
        support and Nikos Kouvaris and Michael Zaks for fruitful
        discussions. LSG thanks support of Humboldt-University at
        Berlin within the framework of German excellence initiative
        (DFG).
\end{acknowledgement}

\FloatBarrier
%\newpage

\textbf{Author contribution statement} \\
\label{sec:acs}
All authors took part in drafting and revising of the manuscript as
well as analysis and interpretation of the data.  SC carried out
simulations and prepared the graphics.
\FloatBarrier

\appendix
\onecolumn
\section{Derivation of the characteristic equation} 
\label{sec:app}

To study the stability of steady states by a characteristic equation,
we introduce the vector $\Pvec(\lambda)$ which is the Laplace
transform of the time dependent deviations $\delta \Pvec(t) =
\Pvec(t)-\Pvec^*$ at a steady state. Then, the linearized version of
\eqref{eq:MFeq_rbnet} reads in Laplace space
\begin{align}
  \lambda\Pvec(\lambda)-\Pvec(t=0) &= \tens{J}(\lambda)
  \Pvec(\lambda),
\end{align}
with the Jacobian $\tens{J}(\lambda)$. Its formal solution is
\begin{align}
  \nonumber \Pvec(\lambda) &= (\lambda\mathbb{1}-\tens{J}(\lambda))^{-1} \Pvec(t=0) \\
  &= \tens{M}^{-1} \Pvec(t=0).
\end{align}

The final value theorem can be used to calculate the steady states of
this system
\begin{align}
  \nonumber \Pvec^* &= \lim_{t\rightarrow\infty} \Pvec(t)=\lim_{\lambda\rightarrow0} \lambda\Pvec(\lambda) \\
  \nonumber &= \lim_{\lambda\rightarrow0} \lambda\tens{M}^{-1} \Pvec(t=0) \\
  &= \lim_{\lambda\rightarrow0} \frac{\lambda}{\det(\tens{M})}
  \adj(\tens{M}) \Pvec(t=0),
\end{align}
with
\begin{align}
  \nonumber &\det(\tens{M}) = \lambda^2-\lambda (1-w_1(\lambda)) \left(\sum_k  \frac{1}{1+\gamma^*_k \langle t\rangle}\left(\frac{\partial\gamma^*_{k}}{\partial P^*_{1,k}}  -\gamma^*_{k}\right)\right) + \\
   &+(1-w_1(\lambda))^2
  \bigg(\gI^*\gII^*-\gI^* \frac{1}{1+\gII^*\langle
    t\rangle}\frac{\partial\gII^*}{\partial \PbetII^*}
  - \gII^* \frac{1}{1+\gI^* \langle t\rangle} \frac{\partial\gI^*}{\partial
    \PbetI^*} \bigg)
  \label{eq:charEqGGp_rbnet}
\end{align}
and
\begin{align}
  \adj(\tens{M}) &= \begin{pmatrix}
    m_{22} & -m_{12} \\
    -m_{21} & m_{11} \\
  \end{pmatrix},
\end{align}
\begin{align*}
  m_{11} &= (1-w_1(\lambda))\left(\frac{1}{1+\gamma^*_{k_1}
    \langle t\rangle} \frac{\partial\gamma^*_{k_1}}{\partial
    P^*(1 | k_1)}-\gamma^*_{k_1}\right)-\lambda, \\ 
m_{12} &= (1-w_1(\lambda)) \frac{1}{1+\gamma^*_{k_1} \langle t\rangle} \frac{\partial\gamma^*_{k_1}}{\partial
  P^*(1 | k_2)},\\ 
m_{21} &= (1-w_1(\lambda)) \frac{1}{1+\gamma^*_{k_2} \langle t\rangle} \frac{\partial\gamma^*_{k_2}}{\partial
  P^*(1 | k_1)},\\ 
m_{22} &= (1-w_1(\lambda))\left(\frac{1}{1+\gamma^*_{k_2}
  \langle t\rangle}\frac{\partial\gamma^*_{k_2}}{\partial
  P^*(1 | k_2)}-\gamma^*_{k_2}\right)-\lambda.
\end{align*}
The final value theorem states that the limit
$\lim_{\lambda\rightarrow0} \lambda\Pvec(\lambda)$ is unique if and
only if the denominator of $\Pvec(\lambda)$ has roots with negative
real parts and not more than one pole at the origin. Thus indicating
bifurcations when one (or several) roots of $\det(\tens{M})$ cross the
imaginary axis. Therefore $\det(\tens{M})=0$ is called the
characteristic equation.

The fact that $w_1(\lambda)$ is the moment generating function of
$w_1(t)$ means that \linebreak $w_1(\lambda)=\sum_{k=0}^\infty
\frac{\langle t^k \rangle}{k!}\lambda^k$ where $\langle t^k \rangle$
is the $k$-th moment of $w_1(t)$. Two typical examples are the pairs
\begin{align}
  w_1(t)=\delta(t-t_w) &\longleftrightarrow w_1(\lambda)=e^{-\lambda t_w} \\
  w_1(t)=\gamma\,e^{-\gamma\,t} &\longleftrightarrow
  w_1(\lambda)=\frac{1}{1+\lambda/\gamma}.
\end{align}

Since $\langle t^0 \rangle = 1$, the term
$(1-w_1(\lambda))=\sum_{k=1}^\infty \frac{\langle t^k
  \rangle}{k!}\lambda^k$ and thus $\adj(\tens{M})$ is of first order
in $\lambda$.  Given this information it is clear that
$\Pvec(\lambda)$ has only one pole of order one at the origin.

\section{Alternative derivation of \eqref{eq:SNEllipseinR_rbnet} using
  the characteristic equation}
  \label{sec:app2}

To look for saddle-node bifurcations the lowest terms in $\lambda$ of
equation \eqref{eq:charEqGGp_rbnet} will be collected. Identifying
$\langle t \rangle = \tau$ results in
\begin{align}
 \nonumber 0 =& 1-\tau\left(\sum_k
\left( \frac{1}{1+\gamma^*_k \tau}\, \frac{\partial\gamma^*_{k}}{\partial P^*(1|k)}
 -\gamma^*_{k}\right)\right)+\\
\nonumber &\tau^2\bigg(\gI^*\gII^*-\gI^*\frac{1}{1+\gII^*\tau} \,\frac{\partial\gII^*}{\partial
  \PbetII}
-\gII^* \frac{1}{1+\gI^* \tau}\, \frac{\partial\gI^*}{\partial
  \PbetI} \bigg)
\end{align}
giving
\begin{align}
  \frac{1}{\tau} = \frac{1}{(1+\gI^*\tau)^2}\,
  \frac{\partial\gI^*}{\partial \PbetI}\, +\,
  \frac{1}{(1+\gII^*\tau)^2}\,\frac{\partial\gII^*}{\partial \PbetII}
  \label{eq:SNEllipseGGP_rbnet}
\end{align}
as the condition for a saddle-node bifurcation.  Application of the
chain rule in the derivatives, i.e. $\partial\gamma/\partial
  P=\partial\gamma/\partial r \, \partial r /\partial P$
finally yields
\begin{align}
  % \nonumber \frac{1}{\tau} &= \frac{1}{(1+\gI^*\tau)^2} \frac{\partial\gI^*}{\partial r^*} \frac{\partial
  %     r^*}{\partial \PbetI}} +
  % \frac{\frac{\partial\gII^*}{\partial r^*} \frac{\partial
  %     r^*}{\partial
  %     \PbetII*}}{(1+\gII^*\tau)^2} \\
  \frac{1}{\tau} &= \frac{k_1 \nu}{\langle
    k\rangle}\frac{1}{(1+\gI^*\tau)^2} \frac{\partial\gI^*}{\partial
      r^*} + \frac{k_2 (1-\nu)}{\langle
    k\rangle}\frac{1}{(1+\gII^*\tau)^2} \frac{\partial\gII^*}{\partial
      r^*}.
\end{align}
This is indeed the same equation as \eqref{eq:SNEllipseinR_rbnet}.
This derivation has the positive side effect that with the aid of the
characteristic equation all other bifurcation scenarios can be
investigated as well.

\twocolumn
\bibliographystyle{iopart-num-SC}
\bibliography{TwoState}

\providecommand{\newblock}{}
\begin{thebibliography}{10}
\expandafter\ifx\csname url\endcsname\relax
  \def\url#1{{\tt #1}}\fi
\expandafter\ifx\csname urlprefix\endcsname\relax\def\urlprefix{URL }\fi
\providecommand{\eprint}[2][]{\url{#2}}
% Bibliography created with iopart-num v2.1
% /biblio/bibtex/contrib/iopart-num

\bibitem{Frantsuzov09}
Frantsuzov P~A, Volk{\'{a}}n-Kacs{\'{o}} S and Jank{\'{o}} B 2009 {\em Phys.
  Rev. Lett.\/} {\bf 103} 207402

\bibitem{Lindner2004}
Lindner B, Garc{\'{i}}a-Ojalvo J, Neiman A and Schimansky-Geier L 2004 {Effects
  of noise in excitable systems}

\bibitem{HuberTsim2005}
Huber D and Tsimring L~S 2005 {\em Phys. Rev. E. Stat. Nonlin. Soft Matter
  Phys.\/} {\bf 71} 36150

\bibitem{PikovskyTsim2001}
Pikovsky A and Tsimring L~S 2001 {\em Phys. Rev. Lett.\/} {\bf 87} 250602

\bibitem{Chen15}
Chen H and Shen C 2015 {\em Phys. A Stat. Mech. its Appl.\/} {\bf 424} 97--104

\bibitem{Dumont2014}
Dumont G, Northoff G and Longtin A 2014 {\em Phys. Rev. E\/} {\bf 90} 12702

\bibitem{PrFaSchiZa07}
Prager T, Falcke M, Schimansky-Geier L and Zaks M~A 2007 {\em Phys. Rev. E\/}
  {\bf 76} 11118

\bibitem{Colaiori15}
Colaiori F, Castellano C, Cuskley C~F, Loreto V, Pugliese M and Tria F 2015
  {\em Phys. Rev. E\/} {\bf 91} 12808

\bibitem{WoodLind06}
Wood K, den Broeck C, Kawai R and Lindenberg K 2006 {\em Phys. Rev. Lett.\/}
  {\bf 96} 145701

\bibitem{WoodLind07}
Wood K, den Broeck C, Kawai R and Lindenberg K 2007 {\em Phys. Rev. E\/} {\bf
  76} 41132

\bibitem{WoodLind06PRE}
Wood K, den Broeck C, Kawai R and Lindenberg K 2006 {\em Phys. Rev. E\/} {\bf
  74} 31113

\bibitem{Prager2003}
Prager T, Naundorf B and Schimansky-Geier L 2003 {Coupled three-state
  oscillators} {\em Phys. A Stat. Mech. its Appl.\/} vol 325 pp 176--185

\bibitem{EscaffLind12}
Escaff D, Harbola U and Lindenberg K 2012 {\em Phys. Rev. E\/} {\bf 86} 11131

\bibitem{Kouvaris2010}
Kouvaris N, M{\"{u}}ller F and Schimansky-Geier L 2010 {\em Phys. Rev. E\/}
  {\bf 82} 61124

\bibitem{Prager2007}
Prager T, Lerch H~P, Schimansky-Geier L and Sch{\"{o}}ll E 2007 {\em J. Phys. A
  Math. Theor.\/} {\bf 40} 11045

\bibitem{Leonhardt2008}
Leonhardt H, Zaks M~A, Falcke M and Schimansky-Geier L 2008 {\em J. Biol.
  Phys.\/} {\bf 34} 521--538

\bibitem{Pinto2014}
Pinto I~L~D, Escaff D, Harbola U, Rosas A and Lindenberg K 2014 {\em Phys. Rev.
  E\/} {\bf 89} 52143

\bibitem{Trimper04}
Trimper S and Zabrocki K 2004 {Memory in diffusive systems}

\bibitem{cox1962renewal}
Cox D~R 1962 {\em {Renewal Theory}\/} Methuen's monographs on applied
  probability and statistics (London: Methuen)

\bibitem{Barrat2004}
Barrat A, Barthelemy M, Satorras R~P and Vespignani A 2004 {\em Proc. Natl.
  Acad. Sci. U. S. A.\/} {\bf 101} 3747--3752

\bibitem{Jiang2007}
Jiang B 2007 {\em Phys. A\/} {\bf 384} 647--655

\bibitem{Mukherjee2003}
Mukherjee G, Sen P, Dasgupta S, Chatterjee A, Sreeram P~A and Manna S~S 2003
  {\em Phys. Rev. E. Stat. Nonlin. Soft Matter Phys.\/} {\bf 67} 36106

\bibitem{Runions2005}
Runions A, Fuhrer M, Lane B, Federl P and Lagan A~G~R 2005 {\em ACM Trans.
  Graph.\/} {\bf 24} 702--711

\bibitem{Ichinomiya2004}
Ichinomiya T 2004 {\em Phys. Rev. E - Stat. Nonlinear, Soft Matter Phys.\/}
  {\bf 70}

\bibitem{Sonnenschein2012}
Sonnenschein B and Schimansky-Geier L 2012 {\em Phys. Rev. E - Stat. Nonlinear,
  Soft Matter Phys.\/} {\bf 85}

\bibitem{Carro2016}
Carro A, Toral R and {San Miguel} M 2016 {\em Sci. Rep.\/} {\bf 6} 24775

\bibitem{Rice1954}
Rice S 1954 {\em {Mathematical analysis of random noise}\/} (Dover)

\bibitem{Hanggi1990}
H{\"{a}}nggi P, Talkner P and Borkovec M 1990 {\em Rev. Mod. Phys.\/} {\bf 62}
  251--341

\bibitem{Kromer2014}
Kromer J~A, Lindner B and Schimansky-Geier L 2014 {\em Phys. Rev. E - Stat.
  Nonlinear, Soft Matter Phys.\/} {\bf 89}

\bibitem{Lambiotte2007}
Lambiotte R 2007 {\em EPL (Europhysics Lett.\/} {\bf 78} 68002

\bibitem{Sonnenschein2013}
Sonnenschein B, Zaks M~A, Neiman A~B and Schimansky-Geier L 2013 {\em Eur.
  Phys. J. Spec. Top.\/} {\bf 222} 2517--2529

\bibitem{phdKouvaris}
Kouvaris N~E 2011 {\em {Study of synchronization in discrete biological
  systems}\/} Ph.D. thesis Aristotle university of Thessaloniki, Department of
  Mathematical, Physical and Computational Sciences

\bibitem{Restrepo2005}
Restrepo J~G, Ott E and Hunt B~R 2005 {\em Phys. Rev. E - Stat. Nonlinear, Soft
  Matter Phys.\/} {\bf 71}

\bibitem{ArenasKuths08}
Arenas A, D{\'{i}}az-Guilera A, Kurths J, Moreno Y and Zhou C 2008 {\em Phys.
  Rep.\/} {\bf 469} 93--153

\bibitem{Boccaletti06}
Boccaletti S, Latora V, Moreno Y, Chavez M and Hwang D~U 2006 {\em Phys.
  Rep.\/} {\bf 424} 175--308

\bibitem{Newman03}
Newman M~E~J 2003 {\em SIAM Rev.\/} {\bf 45} 167--256

\bibitem{Dorogovtsev02}
Dorogovtsev S~N, Goltsev A~V and Mendes J~F~F 2002 {\em Phys. Rev. E\/} {\bf
  66} 16104

\bibitem{Zhou2007}
Zhou H and Lipowsky R 2007 {\em J. Stat. Mech. Theory Exp.\/} {\bf 2007}
  P01009--P01009

\bibitem{Feng2013}
Feng X, Mathoniere C, Jeon I~R, Rouzieres M, Ozarowski A, Mathoni{\`{e}}re C,
  Rouzi{\`{e}}res M, Aubrey M, Gonzalez M, Cl{\'{e}}rac R and Long J 2013 {\em
  J. Am. Chem. Soc.\/} {\bf 135} 15880--15884

\bibitem{Lu2013}
Lu M, Jolly M, Gomoto R, Huang B, Onuchic J and Jacob E~B 2013 {\em J. Phys.
  Chem. B\/} {\bf 117} 13164--13174

\bibitem{Cerna2013}
Cerna R, Leger Y, Paraiso T~K, Wouters M, Genoud F~M, L{\'{e}}ger Y,
  Para{\"{i}}so T~K, Oberli M~T~P and Deveaud B 2013 {\em Nat. Commun.\/} {\bf
  4} 2008

\bibitem{Paraiso2010}
Paraiso T~K, Wouters M, Leger Y, Genoud F~M, Pledran B~D, Para{\"{i}}so T~K,
  L{\'{e}}ger Y and Pl{\'{e}}dran B~D 2010 {\em Nat. Mater.\/} {\bf 9} 655--660

\end{thebibliography}

\end{document}